\pdfoutput=1
\documentclass[a4paper,english,11pt]{amsart}

\usepackage[sc,osf]{mathpazo}
\usepackage[scaled=.95]{helvet} 
\DeclareTextFontCommand{\textsl}{\fontfamily{ppl}\fontshape{sl}\selectfont}
\usepackage[euler-digits]{eulervm}
\let\mathbf\mathbold
\let\oldscshape\scshape
\def\scshape{\oldscshape\lsstyle}
\DeclareTextFontCommand{\textsc}{\oldscshape}

\makeatletter
\def\@captionheadfont#1\@xp{{\scshape #1}\@xp}
\makeatother

\usepackage{babel}
\usepackage{microtype}
\usepackage{mathtools}
\usepackage{amssymb}
\usepackage{amsthm}
\usepackage[foot]{amsaddr}
\usepackage{mathrsfs}
\usepackage{ifthen}
\usepackage{nicefrac}
\usepackage{color}
\usepackage[
pdftitle={Max-Closed Semilinear Constraint Satisfaction},
pdfauthor={Manuel Bodirsky, Marcello Mamino},
pdfkeywords={},
colorlinks=false]{hyperref}

\makeatletter
\def\Hy@Warning#1{}
\makeatother
\makeindex
\makeatletter
\def\@setemails{%
\ifnum\theg@author > 1 
\mbox{{\itshape E-mail addresses}:\space}{\ttfamily\emails}. 
\else 
\mbox{{\itshape E-mail address}:\space}{\ttfamily\emails}. 
\fi%
}
\makeatother
\makeatletter
\def\ps@plain{\ps@empty
  \def\@oddfoot{\normalfont\normalsize \hfil\thepage\hfil}%
  \let\@evenfoot\@oddfoot}
\def\ps@firstpage{\ps@plain
  \def\@oddfoot{\normalfont\normalsize \hfil\thepage\hfil
     \global\topskip\normaltopskip}%
  \let\@evenfoot\@oddfoot
  \def\@oddhead{\@serieslogo\hss}%
  \let\@evenhead\@oddhead 
}
\def\ps@headings{\ps@empty
  \def\@evenhead{%
    \setTrue{runhead}%
    \normalfont\normalsize
    \rlap{\thepage}\hfil
    \textsc{\lsstyle\MakeLowercase{\shortauthors}\hfil}}%
  \def\@oddhead{%
    \setTrue{runhead}%
    \normalfont\normalsize \hfil
    \textsc{\lsstyle\MakeLowercase{\rightmark{}{}}}\hfil\llap{\thepage}}%
  \let\@mkboth\markboth
}
\makeatother
\makeatletter
\let\oldupchars\upchars@
\def\upchars@{\oldupchars\def\-{\U-}}
\makeatother

\def\mathshift{$}
\catcode`$=13
\def${\ifinner\expandafter\mathshift\else\expandafter\myshift\fi}
\def\myshift#1${\raisebox{0ex}[0ex][0ex]{\mathshift#1\mathshift}}
\AtBeginDocument{\catcode`$=13}

\makeatletter
\let\oldsection\section
\def\newsection#1{\oldsection{\lsstyle\ #1}}
\def\newsectionr#1{\oldsection*{\lsstyle #1}}
\def\section{\@ifstar\newsectionr\newsection}
\makeatother

\makeatletter
\newtheorem{ghost@theorem}{}[section]
\def\@maketheorem#1=#2;{
	\newtheorem{#1}[ghost@theorem]{#2}}
\def\maketheorem#1{
	\@for\@x:=#1\do{
		\expandafter\@maketheorem\@x;}}
\makeatother

\def\eqdef{\stackrel{\textsl{\fontsize{6pt}{6pt}\selectfont def}}{=}}

\def\R{\mathbb R}
\def\Q{\mathbb Q}

\def\abs#1{\left|#1\right|}

\def\MMfirstoftwo#1#2{#1}
\def\MMsecondoftwo#1#2{#2}

\def\MMendlist{\MMendlist}
\def\MMuniquetag{\MMuniquetag}

\def\checkempty#1{\MMdocheckempty#1\MMuniquetag\MMendlist}
\def\MMdocheckempty#1#2\MMendlist{\ifx#1\MMuniquetag\expandafter\MMfirstoftwo\else\expandafter\MMsecondoftwo\fi}

\let\n\oldstylenums

\def\mydate#1#2#3{\hbox{\n{#1}$\cdot${\oldscshape #2}$\cdot$\n{#3}}}
\def\new#1#{\MMdonew{#1}}
\def\MMdonew#1#2{\checkempty{#1}{\MMnewtwoparms[#2]{#2}}{\MMnewtwoparms#1{#2}}}
\def\MMnewtwoparms[#1]#2{\textsc{#2}\index{#1}}
\def\-{\nobreakdash-\hspace{0pt}}
\def\U-{\raise0.2ex\hbox{-}}
\def\url#1{\href{#1}{url\nobreakdash---\texttt{#1}}}

\def\mailto#1{\href{mailto:#1}{\texttt{#1}}}
\makeatletter
\def\eatspace#1{#1}
\def\myitem#1{\hfil\break\hbox to 0em{\hss\hbox to 1em{\hfill #1\hfill}. }\def\@currentlabel{#1}\eatspace}
\makeatother

\def\R{\mathbb{R}}
\def\Q{\mathbb{Q}}
\def\P{\ensuremath{\mathsf{P}}}
\def\NP{\ensuremath{\mathsf{NP}}}
\def\coNP{\mathsf{co}\text-\NP}

\def\pmax{\text{\sc max}}
\def\pmin{\text{\sc min}}
\def\vs{\text{\sc s}}
\def\po{\operatorname{po}}
\def\pr{\operatorname{pr}}
\def\vertices{\operatorname{V}}
\def\edges{\operatorname{E}}

\DeclareMathOperator{\Csp}{CSP}
\newcommand{\cproblem}[3]{
\noindent {\bf #1} \\
INSTANCE: #2 \\
QUESTION: #3}

\theoremstyle{definition}
\newtheorem*{theorem*}{Theorem}
\maketheorem{%
	theorem=Theorem,%
	lemma=Lemma,%
	proposition=Proposition,%
	corollary=Corollary,%
	fact=Fact%
}
\maketheorem{%
	definition=Definition,%
	assumption=Assumption,%
	example=Example%
}
\maketheorem{%
	observation=Observation,%
	remark=Remark,%
	claim=Claim,%
	question=Question%
}

\title[Tropically convex constraint satisfaction]{\lsstyle Tropically convex constraint satisfaction}

\date{\mydate{22}{iii}{2017}}

\author[M.\ Bodirsky]{\lsstyle Manuel Bodirsky}
\address{Institut f\"{u}r Algebra, TU Dresden, 01062 Dresden, Germany}
  \email{\mailto{manuel.bodirsky@tu-dresden.de}}
\thanks{Both authors have received funding from the European Research Council (grant agreement number 257039 and 681988), and from the German Research Foundation (DFG, project number 622397).}
\author[M.\ Mamino]{\lsstyle Marcello Mamino}
\email{\mailto{marcello.mamino@tu-dresden.de}}
\thanks{An extended abstract appeared in the proceedings of the
11th International Computer Science Symposium in Russia, CSR 2016.}

\begin{document}

\begin{abstract}
A semilinear relation $S \subseteq {\mathbb Q}^n$ is \emph{max-closed} if
it is preserved by taking the componentwise maximum.  The constraint
satisfaction problem for max-closed semilinear constraints is at least as
hard as determining the winner in Mean Payoff Games, a notorious problem
of open computational complexity.  Mean Payoff Games are known to be
in~$\NP\cap\coNP$, which is not known for max-closed semilinear
constraints.  Semilinear relations that are max-closed and additionally
closed under translations have been called \emph{tropically convex} in the 
literature. One of our main results is a new duality for open tropically
convex relations, which puts the CSP for tropically convex semilinear
constraints in general into~$\NP\cap\coNP$.  This extends the 
corresponding complexity result for
scheduling under and-or precedence constraints,
or equivalently the max-atoms problem.
To this end, we present a characterization of
max-closed semilinear relations in terms of syntactically restricted
first-order logic, and another characterization in terms of a finite set 
of relations $L$ that allow primitive positive definitions of all other
relations in the class.  We also present a subclass of max-closed
constraints where the CSP is in \P; this class generalizes the class of
max-closed constraints over finite domains, and the feasibility problem
for max-closed linear inequalities.  Finally, we show that the class of
max-closed semilinear constraints is \emph{maximal} in the sense that as
soon as a single relation that is not max-closed is added to~$L$, the CSP 
becomes \NP-hard.
\end{abstract}

\maketitle

\vfill
\pagebreak





\section{Introduction}
\noindent
A relation $R \subseteq {\mathbb Q}^n$ is \emph{semilinear} if $R$ has a
first-order definition in $({\mathbb Q};+,\leq,1)$; equivalently, $R$ is a
finite union of finite intersections of (open or closed) linear
half spaces; see Ferrante and Rackoff~\cite{FerranteRackoff}.  In this
article we study the computational complexity of constraint satisfaction
problems with semilinear constraints.  Informally, a constraint
satisfaction problem (CSP) is the problem of deciding whether a given
finite set of constraints has a common solution.  It has been a fruitful
approach to study the  computational complexity of CSPs depending on the
type of constraints allowed in the input. 

Formally, we fix a set $D$, a set of relation symbols $\tau =
\{R_1,R_2\dotsc\}$,
and a $\tau$-structure $\Gamma = (D;R^\Gamma_1,R^\Gamma_2\dotsc)$ where
$R^\Gamma_i \subseteq D^{k_i}$
is a relation over~$D$ of arity~$k_i$. For finite~$\tau$ the computational
problem~$\Csp(\Gamma)$ is defined as follows.
\vskip.5\baselineskip
\cproblem{$\Csp(\Gamma)$}
{a finite set of formal variables $x_1\dotsc x_n$,
and a finite set of expressions of the form $R(x_{i_1}\dotsc x_{i_k})$
with
$R \in \tau$}
{is there an assignment $x^s_1\dotsc x^s_n\in D$ such
that $(x^s_{i_1}\dotsc x^s_{i_k}) \in R^\Gamma$ for all
constraints of the form $R(x_{i_1},\dots,x_{i_k})$ in the input?}
\vskip.5\baselineskip
When the domain of $\Gamma$ is the set of rational numbers $\mathbb Q$,
and all relations of $\Gamma$ are semilinear, we say that $\Gamma$ is
semilinear. 
We adopt the convention that all semilinear relations have rational
coefficients. 
For relations of this kind and the questions studied here, whether we work with $D =
{\mathbb Q}$ or with $D = {\mathbb R}$ does not play any role. 

It is possible, and sometimes essential, to also define~$\Csp(\Gamma)$
when the signature~$\tau$ is infinite.  However, in this situation it is
important to discuss how the symbols from~$\tau$ are represented in the
input of the~CSP. In our context, $\Gamma$ is semilinear, it is natural to
assume that the relations are given as quantifier-free formulas in
disjunctive normal form where the coefficients are represented in binary
(by the mentioned result of Ferrante and Rackoff~\cite{FerranteRackoff},
every relation with a first-order definition over $({\mathbb Q};+,\leq,1)$
has a definition of this form).  However, there are other natural
representations, and we will for infinite languages always discuss how the
relations are given. 



A famous example of a computational problem that can be formulated as
$\Csp(\Gamma)$ for a semilinear structure~$\Gamma$ is the feasibility
problem for linear programming, which is simply the CSP for the
structure~$\Gamma$ with domain~$\mathbb Q$ that contains a relation for
every linear inequality, that is, all relations of the form $\big
\{(x_1,\dots,x_k) \; | \; c_1 x_k + \cdots + c_k x_k \leq c_0 \big \}$ for
coefficients $c_0,c_1,\dots,c_k \in \mathbb Q$.    Here, it is natural to
assume that the relations in the input are represented via the
inequalities that define them, and that the coefficients are represented
in binary.  This CSP is well-known to be in $\P$~\cite{Khachiyan}. It
follows that the CSP for \emph{any} semilinear~$\Gamma$ (represented, say,
in~\textsc{dnf}) is in \NP, because we can non-deterministically select a
disjunct from the representation of each of the given constraints, and
then verify in polynomial time whether the obtained set of linear
inequalities is satisfiable.

We would like to systematically study the computational complexity of
$\Csp(\Gamma)$ for all semilinear structures $\Gamma$.  This is a very
ambitious goal.
Several partial results are
known~\cite{Semilinear,Essentially-convex,BodMarMot,JonssonLoeoew,JonssonThapper}.
Let us also mention that it is easy to find for every structure $\Delta$ with a finite domain
a semilinear structure $\Gamma$ so that
$\Csp(\Delta)$ and $\Csp(\Gamma)$ are the same computational problem (in
the sense that they have the same satisfiable instances). But already the
complexity classification of CSPs for finite structures is
open~\cite{FederVardi}.

Even worse, there are concrete semilinear structures whose CSP has an open
computational complexity.  An important example of this type is the
max-atoms problem~\cite{Max-atoms}, which is the CSP for the semilinear
structure $\Gamma$ that contains all ternary relations of the form
\begin{align*}
	M_c \eqdef \big\{(x_1,x_2,x_3) \;{\big|}\; x_1 + c \leq
	\max(x_2,x_3)\big\}
\end{align*}
where $c \in \mathbb Q$ is represented in binary.  It is an open problem
whether the max atoms problem is in \P, but it is known to be
polynomial-time equivalent to determining the winner in mean payoff games
(M\"ohring, Skutella, and Stork~\cite{and-or-scheduling}), which is known
to be in~$\NP\cap\coNP$.  Note that here the assumption that $c$ is
represented in binary is important: when $c$ is represented in unary, or
when we drop all but a finite number of relations in $\Gamma$, the
resulting problem is known to be in~\P.

An important tool to study the computational complexity of $\Csp(\Gamma)$
is the concept of \emph{primitive positive definability}.  A primitive
positive formula is a first-order formula of the form $\exists
x_1,\dots,x_n (\psi_1 \wedge \cdots \wedge \psi_m)$ where $\psi$ are
atomic formulas (in primitive positive formulas, no disjunction, negation,
or universal quantification is allowed).  Jeavons, Cohen, and
Gyssens~\cite{JeavonsClosure} showed that the CSP for expansions of $\Gamma$
by finitely many primitive positive definable relations is polynomial-time
reducible to $\Csp(\Gamma)$.  Let us mention that difficult problems in
real algebraic geometry are about primitive positive definability: the
conjecture of Helton and Nie~\cite{HeltonNieMP10} for instance can be
phrased as ``every convex semialgebraic set has a primitive positive
definition over the solution spaces of semidefinite programs''. 

Primitive positive definability in $\Gamma$ can be studied using the
\emph{polymorphisms} of $\Gamma$, which are a multi-variate generalization
of the endomorphisms of $\Gamma$.  We say that $f \colon \Gamma^k \to
\Gamma$ is a polymorphism of a $\tau$-structure $\Gamma$ if \begin{align*}
\big (f(a^1_1,\dots,a^k_1),\dots,f(a^1_m,\dots,f^k_m) \big ) \in R^\Gamma
\end{align*} for all $R \in \tau$ and
$(a^1_1,\dots,a^1_m),\dots,(a^k_1,\dots,a^k_m) \in R^\Gamma$.  For finite
structures $\Gamma$, a relation $R$ is primitive positive definable in
$\Gamma$ \emph{if and only if} $R$ is preserved by all polymorphisms of
$\Gamma$.  And indeed, the \emph{tractability conjecture} of Bulatov,
Jeavons, and Krokhin~\cite{JBK} in a reformulation due to Kozik and
Barto~\cite{Cyclic} states that $\Csp(\Gamma)$ is in~\P\ if and only if
$\Gamma$ has a polymorphism $f$ which is \emph{cyclic}, this is, has arity
$n \geq 2$ and satisfies $\forall x_1,\dots,x_n \; f(x_1,\dots,x_n) =
f(x_2,\dots,x_n,x_1)$.

Polymorphisms are also relevant when $\Gamma$ is infinite. For example, a
semilinear relation is convex if and only if it has the cyclic polymorphism
$(x,y) \mapsto (x+y)/2$.  The CSP for such relations is the feasibility
problem for strict and non-strict linear inequalities, and it is
well-known that this problem is in~\P.  When $\Gamma$ has a cyclic
polymorphism, and assuming the tractability conjecture, $\Gamma$ cannot
interpret\footnote{Interpretations in the sense of model theory; we refer
to Hodges~\cite{Hodges} since we do not need this concept further.}
primitively positively any hard finite-domain CSP, which is the standard
way of proving that a CSP is NP-hard.

A fundamental cyclic operation is the maximum operation, given by
$(x,y) \mapsto \max(x,y)$.  The constraints for the max-atoms problem are
examples of semilinear relations that are not convex, but that have $\max$
as a polymorphism; we also say that they are \emph{max-closed}.  When a
\emph{finite} structure is max-closed, with respect to some ordering of
the domain, then the CSP for this structure is known to be in
$\P$~\cite{Ordered}.  The complexity of the CSP for max-closed
\emph{semilinear} constraints, on the other hand, is open.  Since this
problem is more general than the max atoms problem, it is at least as hard
as determining the winner of mean payoff games. But unlike the max-atoms
problem, it is not known whether the CSP for max-closed semilinear
constraints is in~$\coNP$.


\section{Results}
\noindent
We show that the CSP for semilinear max-closed relations that are
\emph{translation-in\babelhyphen{soft}va\babelhyphen{soft}ri\babelhyphen{soft}ant}, that is,
have the polymorphism $x \mapsto x+c$ for all $c \in \mathbb Q$, is in
$\NP\cap\coNP$ (Section~\ref{sec-csp}).
Such relations have been called \emph{tropically convex}
in the literature~\cite{DevelinSturmfels}\hbox to 0pt{.\hss}\footnote{The original definition
of tropical convexity is for the dual situation, considering $\min$
instead of $\max$.}
This class is a non trivial extension of the max-atoms problem
(for instance it contains relations such as~$x \leq (y+z)/2$), and it is
not covered by the known reduction
to mean payoff
games~\cite{and-or-scheduling,AkianGaubertGuterman,AtseriasManeva}.
Indeed, it is open whether the CSP for tropically convex semilinear
relations can be reduced
to mean payoff games
(in fact, Zwick and Paterson~\cite{ZwickPaterson} believe that mean payoff
games
are ``strictly easier'' than simple stochastic games, which reduce to our
problem via the results presented in Section~\ref{sect:duality}).
The containment
in $\NP\cap\coNP$ can be slightly extended to the CSP for the structure
that includes additionally all relations~$x=c$ for $c \in {\mathbb Q}$
(represented in binary).
It follows from our results
(Corollary~\ref{th-smallest-class}) that
the class of semilinear tropically convex sets is the smallest
class of
semilinear sets that has the same polymorphisms as the max atoms
constraints $x\le\max(y,z)+c$ with $c\in\Q$.

In our proof, we first present
a characterization of max-closed semilinear relations
in terms of syntactically restricted first-order logic
(Section~\ref{sect:syntax}). We show that a semilinear relation is
max-closed
if and only if it can be defined by a \emph{semilinear Horn formula},
which we define as a finite conjunction of \emph{semilinear Horn clauses},
this is,
finite disjunctions of the form 
\begin{align*}
\bigvee_{i=1}^m \bar a_i^{\top} \bar x \succ_i c_i 
\end{align*}
where
\begin{enumerate}
\item $\bar a_1\dotsc\bar a_m \in {\mathbb Q}^n$
and there is a $k \leq n$ such that $\bar a_{i,j} \geq 0$ for all $i$ and
$j\neq k$,
\item $\bar x = (x_1\dotsc x_n)$ is a vector of variables,
\item $\succ_i \, \in \, \{\geq,>\}$ are strict or non-strict
inequalities,
and
\item $c_1\dotsc c_m \in \mathbb Q$ are coefficients.
\end{enumerate}

\begin{example}\label{ex1}
The ternary relation $M_c$ from the max-atoms 
problem can be defined by the semilinear Horn clause
$x_2 - x_1 \geq c \vee x_3 - x_1 \geq c$. 
\end{example}

\begin{example}\label{ex2}
A linear inequality $a_1 x_1 + \cdots + a_n x_n \geq c$ is max-closed if and only if at most one of $a_1\dots a_n$ is negative. The relations defined by such formulas are in general \emph{not} tropically convex; consider for example the relation defined by $-x_1+x_2+x_3 \geq 0$. 
\end{example}

\begin{example}\label{ex3}
Conjunctions of implications of the form
\begin{align}
x_1 \leq c_1 \wedge \cdots \wedge x_n \leq c_n \quad{\Rightarrow}\quad x_i < c_0 \label{eq:jeavons}
\end{align}
are max-closed 
since such an implication is equivalent to the semilinear Horn clause
\begin{align*}
(-x_i > -c_0) \vee \bigvee_i x_i>c_i \; .
\end{align*} 
\end{example}
It has been shown by Jeavons and Cooper~\cite{Ordered} that over finite
ordered domains, a relation is max-closed\footnote{Also the results in
Jeavons and Cooper~\cite{Ordered} have been formulated in the
dual situation for $\min$ instead of $\max$.} if and only if it
can be defined by finite conjunctions of implications of the
form~(\ref{eq:jeavons}). Over infinite domains, this is no longer
true, as demonstrated by the relations in Example~\ref{ex1} and
Example~\ref{ex2}.

We also show that the classes~$\mathcal C$
of max-closed semilinear relations and~$\mathcal C_t$ of tropically convex
semilinear relations are 
\emph{finitely related} in the sense of universal algebra,
this is, there exists a finite subset~$\Gamma_0$ of~$\mathcal C$
(resp.~$\Gamma_t$ of~$\mathcal C_t$)
that can primitively positively define all other relations in~$\mathcal C$
(resp.~$\mathcal C_t$).
For instance, we show
that all relations in~$\mathcal C$ have a primitive positive definition in 
\begin{align*}
\Gamma_0 = ({\mathbb Q};<,1,-1,S_1,S_2,M_0)
\end{align*}
where
\begin{align*}
S_1 = \big \{(x,y) \;{\big|}\; 2x \leq y \big\}
&\quad \quad \quad \quad
S_2 = \big \{(x,y,z) \;{\big|}\; x \leq y + z\big \}\\
M_0 = \big \{(x,&y,z) \;{\big|}\; x \leq y \vee x \leq z \big\}
\end{align*}
Note that any other structure~$\Gamma_1$ with finite relational signature
and this property has a polynomial-time equivalent CSP. We show that the
primitive positive formulas can even be computed efficiently from a given
semilinear Horn formula $\phi$, and have linear size in the representation
size of~$\phi$. 
\begin{figure}
\begin{center}
\includegraphics[width=.8\textwidth]{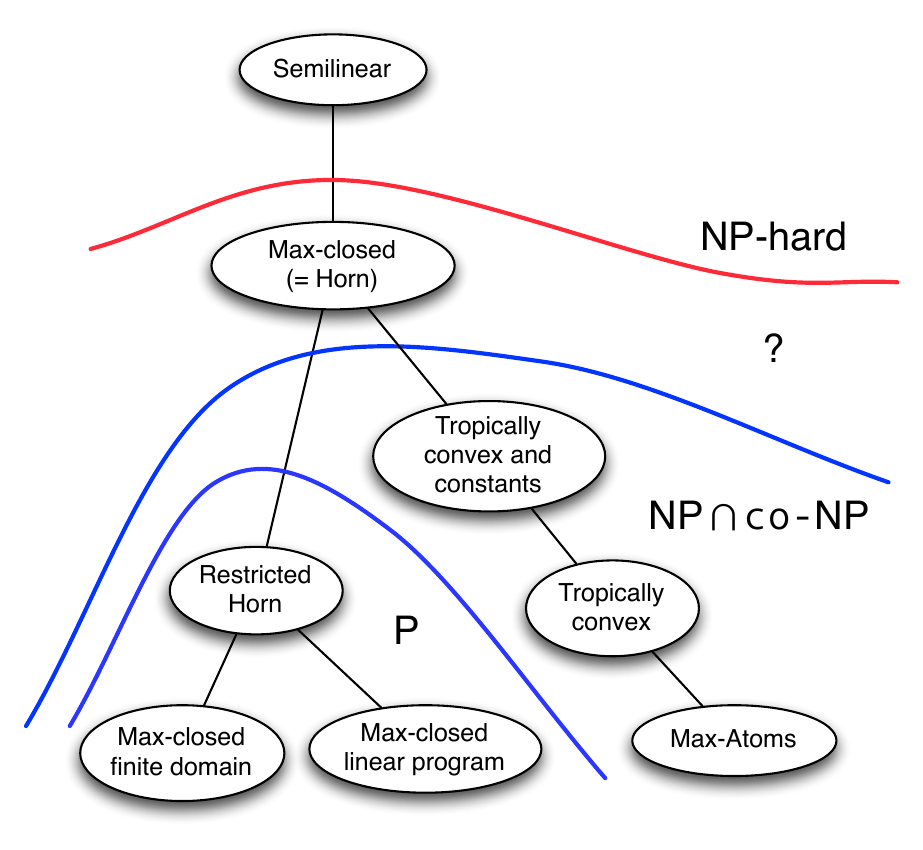}
\end{center}
\caption{An overview of constraint languages relevant for this work
($\P$ membership for finite domain max-closed CSPs is in~\cite{Ordered},
$\NP\cap\coNP$ membership of max-atoms is in~\cite{Max-atoms}).}
\label{fig:pict}
\end{figure}

Our proof of the containment in~$\NP\cap\coNP$ is based on a duality for
open tropically convex semilinear sets, which extends a duality that has
been observed for the max-atoms problem in~\cite{GrigorievPodolskii}.  To
prove the duality, we translate instances of our problem into a condition
on associated mean payoff stochastic games (also called limiting average
payoff stochastic games; see~\cite{Filar-Vrieze} for a general reference),
and then exploit the symmetry implicit in the definition of such games.
The connections between the max-atoms problem, mean payoff deterministic
games, and tropical polytopes have been explored extensively in the
computer science literature. However, the theory of stochastic games
introduces profound changes over the deterministic setting, and employs
non-trivial new techniques. Even though this field is active since the
70s, to the best of our knowledge, no application of its results to
semilinear feasibility problems has been published yet. Note that to solve
mean payoff stochastic games is in~$\NP\cap\coNP$, as a consequence of the
existence of optimal positional strategies; see \cite{Gillette} and
\cite{LiggettLippman} (for a general reference on the computational
complexity of stochastic games, see \cite{andersson-miltersen}).  However,
we cannot use this fact directly, because stochastic games only relate to
a subset of tropically convex sets.  We conclude our argument combining
the duality with semilinear geometry techniques.

Interestingly, at several places in our proofs, we need to
replace $\Q$ with appropriate non-Archimedean structures, for instance in
the proof of the syntactic characterization of max-closed semi-linear
relations, in order to deduce the general case from the easier situation
for closed semilinear relations.  But also for representing the winning
strategies for stochastic games, and for using the duality for closed
tropically convex relations to solve the CSP for tropically convex
constraints, we work with explicit non-standard models (for another
application of non-standard models to CSPs see for instance~\cite{BodMarMot}).

Our next result is the identification of a class of max-closed semilinear
relations whose CSP can be solved in polynomial time
(Section~\ref{sect:poly}). This class consists of the semilinear relations
that can be defined by \emph{restricted semilinear Horn formulas}, which
are conjunctions of semilinear Horn clauses satisfying the additional
condition that there are $k \leq n$ and $l \leq m$ such that $({\bar
a_i})_j \geq 0$ for all $i \in \{1,\dots,m\} \setminus \{l\}$ and $j \in
\{1,\dots,n\} \setminus \{k\}$. An example of a semilinear Horn clause
which is not restricted Horn is $x_1 < x_2 \vee x_1 < x_3$.

Finally, we show that the class of max-closed semilinear constraints is
\emph{maximal} in the sense that for every relation $R$ that
is not max-closed, the problem $\Csp(\Gamma_0,R)$ is \NP-hard
(Section~\ref{sect:maximal}).  Figure~\ref{fig:pict} gives an overview
over our results.  Of course, all our results apply dually to min-closed
semilinear relations as well.

\section{A Syntactic Characterization of Max-closure}
\label{sect:syntax}
\label{sect-syntax}
\noindent
This section aims to offer a syntactic characterization of semilinear
max-closed and semilinear tropically convex relations over~$\Q$. In this section, the
letter~$F$ will denote an ordered field: for technical reasons we need to
work in this slightly more general setting.

\begin{definition}
A semilinear Horn clause is called~\emph{closed}
if all inequalities are non-strict.
We say that $X\subset F^n$ is a \emph{basic max-closed set} if it is the
graph of a semilinear Horn clause, i.e.\ there is~$k\in\{1\dotsc n\}$
such that $X$ can be written as a finite union
\[
  X = \bigcup_i
    \big \{ (x_1\dotsc x_n) \;{\big|}\; a_{i,1}x_1 + \dotsb + a_{i,n}x_n \succ_i c_i \big \}
\]
where ${\succ_i}$ can be either~${>}$ or~${\ge}$, and $a_{i,j}\ge 0$ for
all~$i$ and all~$j\neq k$.  We say that $X$ is \emph{basic closed
max-closed} if it is the graph of a closed semilinear Horn clause.
\end{definition}

\begin{theorem}\label{th-normal-form}
\begin{enumerate}
\item Let $X\subset F^n$ be a semilinear set. Then $X$ is max-closed
if and only if it is a finite intersection of basic max-closed sets.
\item Let $X\subset F^n$ be a closed semilinear set. Then $X$ is
max-closed
if and only if it is a finite intersection of basic closed max-closed
sets.
\item $X\subset\Q^n$ is primitive positive definable in
$\Gamma_0 = ({\mathbb Q};<,1,-1,S_1,S_2,M_0)$
with
\begin{align*}
S_1 = \big \{(x,y) \;{\big|}\; 2x \leq y \big\}
&\quad \quad \quad \quad 
S_2 = \big \{(x,y,z) \;{\big|}\; x \leq y + z\big \}\\
M_0 = \big \{(x,&y,z) \;{\big|}\; x \leq y \vee x \leq z \big\}
\end{align*}
if and only if $X$ is semilinear and max-closed.
\item $X\subset\Q^n$ is primitive positive definable in
$\Gamma_0' = ({\mathbb Q};1,-1,S_1,S_2,M_0)$
if and only if $X$ is semilinear closed and max-closed.
\end{enumerate}
\end{theorem}

\begin{definition}
We say that $X\subset F^n$ is a \emph{basic tropically convex set} if
there is~$k\in\{1\dotsc n\}$
such that $X$ can be written as a finite union
\[
  X = \bigcup_i
    \big \{ (x_1\dotsc x_n) \;{\big|}\; a_{i,1}x_1 + \dotsb + a_{i,n}x_n \succ_i c_i \big \}
\]
where ${\succ_i}$ can be either~${>}$ or~${\ge}$,
and $a_{i,j}\ge 0$ for all~$i$ and all~$j\neq k$, and, moreover
\[
\sum_j a_{i,j} = 0
\]
for all $i$.
\end{definition}

\begin{theorem}\label{th-syntax-tropical}
\begin{enumerate}
\item Let $X\subset F^n$ be a semilinear set. Then $X$ is tropically convex
if and only if it is a finite intersection of basic tropically convex
sets.
\item $X\subset\Q^n$ is primitive positive definable in
$
\Gamma_t = ({\mathbb Q};<,T_1,T_{-1},S_3,M_0)
$ where
\[
T_{\pm1} = \big\{(x,y) \;{\big|}\; x \le y\pm1 \big\}
\quad\quad
S_3 = \left\{(x,y,z) \;{\Big|}\; x \le \frac{y+z}{2} \right\}
\]
if and only if $X$ is semilinear and tropically convex.
\end{enumerate}
\end{theorem}

\begin{corollary}\label{th-smallest-class}
A semilinear set $X \subset \Q^n$ is tropically convex if and only if it is
preserved by every polymorphism that preserves the max-atoms language
(i.e.\ all sets of the form~$\big\{(x,y,z) \;\mid\; x \le \max(y,z)+c\big\}$
for $c \in \Q$).
\end{corollary}

\begin{proof}
Translations and maximum are polymorphisms of max-atoms, so one 
direction is trivial. For the converse, by Theorem~\ref{th-syntax-tropical}(2), it suffices
to prove that the relations $<$, $T_1$, $T_{-1}$, $S_3$, and~$M_0$ are 
preserved by all polymorphisms of max-atoms.
This is immediate for $T_1$, $T_{-1}$, and~$M_0$ since they have primitive positive
definitions over max-atoms. The relation~$<$, on the other hand, is an
monotone union of max-atoms constraints
\[{<} = \big\{ (x,y) \;\mid\; x < y \big\} = \bigcup_{c \in \Q, c>0} \big\{ (x,y) \;\mid\; x \le y - c \big\} \]
and $S_3$ is an intersection of max-atoms constraints
\[S_3 = \left\{ (x,y,z) \;\mid\; x <= \frac{y+z}{2} \right\} = \bigcap_{c \in \Q} \big\{ (x,y,z) \;\mid\; z
\le \max(y+c, z-c) \big\}\]
It's well known that monotone unions and arbitrary intersections of primitive
positive definable sets are preserved by polymorphisms~\cite{PoeschelLoc}.\qed
\end{proof}

The following observation is important in view of its implications on the
complexity of the constraint satisfaction problems for max-closed
and tropically convex sets. The reader will
recognize that it follows straightforwardly from the proofs of
Theorem~\ref{th-normal-form} and Theorem~\ref{th-syntax-tropical}.

\begin{observation}\label{th-concise}
Given a max-closed (resp.\ tropically convex) set~$X$ written as a finite
intersection of basic max-closed (resp.\ basic tropically convex) sets
with the constants represented in binary, we can compute in polynomial
time a primitive positive definition of~$X$ in the structure~$\Gamma_0$
(resp.\ $\Gamma_t$).
\end{observation}

In this section, as well as in Section~\ref{sect-tconv}, we will make use
of a few basic facts about semilinear geometry. Semilinear sets, as
defined in the introduction, are finite Boolean combinations of half
spaces in~$\Q^n$. We already mentioned that this setting can be extended
without difficulty to an ordered field~$F$. An example of ordered field
that we will need is the field of formal Laurent series $F((\epsilon))$,
whose elements are the series of the form
\[
	\sum_{i=-\infty}^\infty a_i\epsilon^i
\]
where $\epsilon$ is a formal variable, the coefficients~$a_i$ are in~$F$,
and the set of integers~$i$ such that $a_i\neq0$ is bounded from below. The
field~$F((\epsilon))$
can be seen as the field of fractions of the ring of formal power
series~$F[[\epsilon]]$, whose elements are
\[
	\sum_{i=0}^\infty a_i\epsilon^i
\]
On both these structures, we give the lexicographic order with
$\epsilon^i\ll\epsilon^j$ for~$i>j$, the choice of~$\epsilon$ as the name
of the formal variable is indeed meant to be a reminder of this. The open
convex semilinear subsets of~$F^n$ are precisely the finite intersections
of open half spaces (see~\cite[Corollary~4.9]{AnRuVe06} and also~\cite{Scow09}).

Still further, one can
consider finite Boolean combinations of half spaces in~$V^n$, where $V$ is
an ordered $F$-vector space. By a half space in~$V^n$ here we mean a set
defined by~$\{\bar x\in V^n\;|\;\bar a^\top \bar x \succ c\}$, where $\bar a\in
F^n$ and~$c\in V$. From this point of view, one can prove that the
semilinear sets are precisely those first order definable in the
structure~$\Gamma_V=(V;<,+,\alpha\cdot)_{\alpha\in F}$, where $\alpha\cdot$
represents the function mapping $x\in V$ to~$\alpha x$. Now, $\Gamma_V$ is
a so called o-minimal structure, see~\cite{VanDenDries} as a general
reference, and, as all o-minimal structures expanding an ordered group, it
has definable Skolem functions (definable choice). Namely, given a parametric existential
formula $\exists x \phi(x,\bar p)$ which is true for all~$\bar p\in V^n$,
there is a semilinear function $x = x(\bar p)$ such that
$\phi(x(\bar p),\bar p)$ holds for all~$\bar p\in V^n$. We will make extensive use of
this principle.

We begin with the proof of Theorem~\ref{th-normal-form}. It is
convenient to prove the four points in the order (2)--(1)--(3)--(4).
Then we will use Theorem~\ref{th-normal-form} to deduce
Theorem~\ref{th-syntax-tropical}.

\begin{claim}\label{th-convex-open}
Let $X\subset F^n$ be an open semilinear set, then $X$ is a finite
union of convex open semilinear sets.
\end{claim}
\begin{proof}
A semilinear set~$Y\subset F^n$ is said to be \emph{relatively open} if $Y$
is open in the affine subspace of~$ F^n$ generated by~$Y$. All
semilinear sets are finite unions of relatively open convex
semilinear sets (e.g.\ \cite[Chapter~1 Corollary~7.8]{VanDenDries}). It therefore suffices to show that given a relatively
open convex semilinear $Y\subset X$, there is an open convex
semilinear set~$Z$ such that $Y\subset Z\subset X$.
Let $A$ be the affine space generated by~$Y$. Then $Y$ can be written
in the form
\[
Y = \big \{ x\in A \;{\big|}\; f_1(x)>0 \wedge \dotsb \wedge f_p(x)>0 \big\}
\]
for some affine $f_1\dotsc f_p$.  Without loss of generality, we may
assume that the projection of~$A$ onto the first~$d\eqdef\dim(A)$ coordinates is
one-to-one. Consider the
function~$\delta\colon F^{>0}\to F^{\ge0}\cup \{+\infty\}$ mapping~$\epsilon$ to the
largest~$\delta(\epsilon)$ such that for all~$x\in Y$ we have
\[
f_1(x)\ge\epsilon \wedge \dotsb \wedge f_p(x)\ge\epsilon \quad \Rightarrow\quad
x+B_{\delta(\epsilon)} \subset X
\]
where
\[
	B_\alpha = \big\{ y\in F^n \;{\big|}\; y_1=\dotsb=y_d=0 \;{\wedge}\; y_{d+1}\dotsc
y_n\in{(}-\alpha,\alpha{)} \big\}
\]
First we claim that $\delta(\epsilon)$ is well defined and positive for
all~$\epsilon>0$. Well definedness is immediate. Suppose that for
some~$\epsilon$ we have~$\delta(\epsilon)=0$. This means that for
all~$\delta'>0$ we can find an~$x'(\delta')\in Y$ such that
\[
f_1(x'(\delta'))\ge\epsilon \wedge \dotsb \wedge
f_p(x'(\delta'))\ge\epsilon \quad {\wedge}\quad
x'(\delta')+B_{\delta'} \nsubseteq X
\]
By definable choice we can assume that the function~$\delta'\mapsto
x'(\delta')$ is semilinear. Consider the limit
\[
x'' = \lim_{\delta'\to 0+} x'(\delta')
\]
Clearly, for all $\delta''>0$
\[
f_1(x'')\ge\epsilon \wedge \dotsb \wedge
f_p(x'')\ge\epsilon \quad{\wedge}\quad
x''+B_{\delta''} \nsubseteq X
\]
contradicting the fact that $X$ is open. Now,
$\epsilon\mapsto\delta(\epsilon)$ is first-order definable from
semilinear data, hence it is semilinear, and we know that it must be
increasing and map positive elements to positive elements. It follows that
$\delta(\epsilon) \le a \min(\epsilon,b)$ for some positive~$a$ and~$b$.
Hence
\[
Z = \bigcup_{x\in Y} x+B_{a\min(\epsilon,b)} \subset X
\]
It is easy to check that $Z$ is open and convex.
\end{proof}

\begin{figure}
\begin{center}
\def\svgwidth{\textwidth}
\input{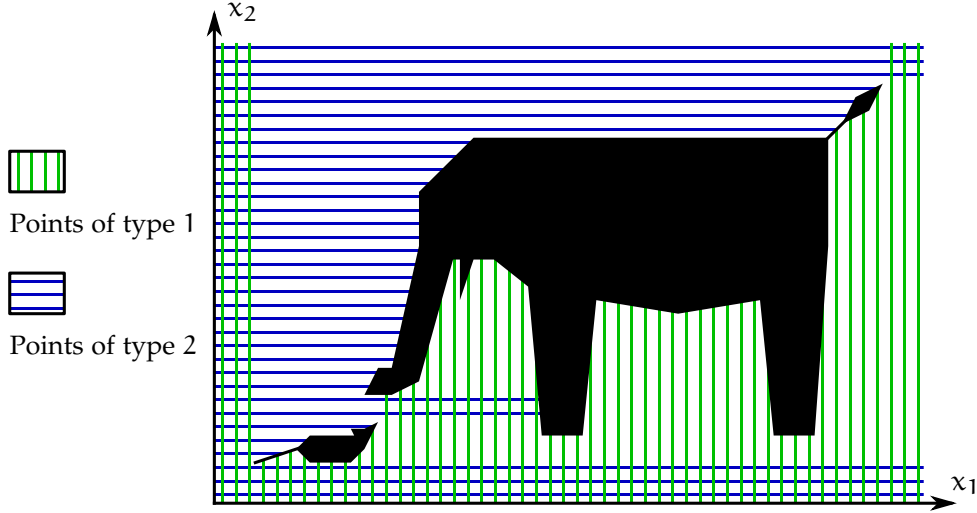}
\end{center}
\caption{A max-closed set in $\Q^2$}
\label{fig-ele}
\end{figure}

\begin{definition}
We say that $x\in X\subset F^n$ is \emph{of type~$k$ in~$X$}, with $k\in\{1\dotsc n\}$,
if $x-Q_k\eqdef\{x-y\;{|}\;y\in Q_k\}\subset X$ where
\[
  Q_k = \big\{ y\in( F^{\ge0})^n \;{\big|}\; y_k = 0 \big\}
\]
\end{definition}
Observe that if $X$ is the complement of a max-closed set, then every
point of~$X$ is of type~$k$ in $X$ for at least one~$k$.
Figure~\ref{fig-ele} displays an example of semilinear max-closed set, with points of
its complement marked according to their type (notice that max-closed sets
need to be neither convex nor connected).

\begin{claim}\label{th-type-k}
Let $X$ be an open semilinear subset of~$ F^n$. Then the
set~$X_k$ of all points of type~$k$ in~$X$ is an open semilinear set.
\end{claim}
\begin{proof}
The set~$X_k$ is clearly semilinear, hence all we need to prove is
that it is open. Pick $x$ on the boundary of~$X_k$, it suffices to prove
that $x\notin X_k$. Since $x$ is on the boundary
\[
{\forall \epsilon>0} \quad {\exists\, \delta(\epsilon) \in F^n} \quad
{\abs{\delta(\epsilon)}_\infty < \epsilon} \quad {\wedge}\quad {\delta(\epsilon)+x \notin X_k}
\]
where $\abs{\,\cdot\,}_\infty$ denotes the maximum norm,
and expanding the definition of~$X_k$
\[
{\forall \epsilon>0} \quad {\exists\, \delta(\epsilon) \in F^n} \quad
{\abs{\delta(\epsilon)}_\infty < \epsilon} \quad {\wedge} \quad
{\exists\, y(\epsilon)\in Q_k} \quad {\delta(\epsilon)+x-y(\epsilon) \notin X}
\]
Now, by definable choice, we can assume that the
function~$\epsilon\mapsto y(\epsilon)$ is semilinear. Hence the
limit
\[
y_0 = \lim_{\epsilon\to0+} y(\epsilon)
\]
exists. Since $Q_k$ is closed, we have $y_0\in Q_k$, and, since $X$ is open,
we have $x-y_0\notin X$, therefore $x\notin X_k$.
\end{proof}

\begin{proof}[Proof of Theorem~\ref{th-normal-form} (2)]
The {\it if\/}~part is immediate, hence we concentrate on the~{\it only if}.
Let $\bar X$ be~$ F^n\setminus X$.
Consider the subsets
$\bar X_1\dotsc \bar X_n$ of the points in~$\bar X$ of type~$1\dotsc n$
respectively.
Since $X$ is max-closed, each point
of~$\bar X$ is of type~$k$ for some~$k\in\{1\dotsc n\}$, hence
\[
\bar X = \bar X_1 \cup \dotsb \cup \bar X_n
\]
moreover, by Claim~\ref{th-type-k}, each of these sets is open and
semilinear. Applying Claim~\ref{th-convex-open}, we can further
split each of the sets~$\bar X_i$ into a finite union
\[
\bar X_i = \bar X_{i,1} \cup \dotsb \cup \bar X_{i,m_i}
\]
of convex open semilinear sets. Now, by definition, $X_i = X_i -
Q_i$, hence
\[
\bar X_i = \tilde X_{i,1} \cup \dotsb \cup \tilde X_{i,m_i}
\]
where $\tilde X_{i,j} = \bar X_{i,j}-Q_i$. We claim that each of the
sets~$X_{i,j} =  F^n \setminus \tilde X_{i,j}$ is definable by a
semilinear Horn clause: this is enough to conclude because, by what was
said, $X$ is the finite intersection
\[
X = \bigcap_{i,j} X_{i,j}
\]
It is easy to check that the
sets~$\tilde X_{i,j}$ are open and convex. Therefore each of them is a
finite intersection
\[
  \tilde X_{i,j} = \bigcap_k
    \big\{ (x_1\dotsc x_n) \;{\big|}\; a_{i,j,k,1}x_1 + \dotsb + a_{i,j,k,n}x_n
    < c_{i,j,k} \big\}
\]
in other words
\[
  X_{i,j} = \bigcup_k
    \big\{ (x_1\dotsc x_n) \;{\big|}\; a_{i,j,k,1}x_1 + \dotsb + a_{i,j,k,n}x_n
    \ge c_{i,j,k} \big\}
\]
It remains to check the condition on the coefficients~$a_{i,j,k,l}$,
namely we claim that $a_{i,j,k,l}\ge 0$ for all~$l\neq i$. For a
contradiction, suppose that $a_{i,j,k,l}<0$ with~$l\neq i$. Then pick a
point~$(x_1\dotsc x_n)\in \tilde X_{i,j}$. Then
\[
(x_1\dotsc x_n) - (0\dotsc \overset{l}{N} \dotsc 0) \notin \tilde X_{i,j}
\]
for $N$ larger than
\[
N' = \frac{a_{i,j,k,1}x_1 + \dotsb + a_{i,j,k,n}x_n}{-a_{i,j,k,l}}
\]
contradicting the construction of~$\tilde X_{i,j}$.
\end{proof}

\begin{proof}[Proof of Theorem~\ref{th-normal-form}---{\rm(}2\/{\rm)}$\Rightarrow${\rm(}1\/{\rm)}]
Our strategy is the following. We apply~(2) to the field of formal Laurent
series~$F((\epsilon))$ with coefficients in~$F$.
Each semilinear set~$X\subset F^n$
has a unique extension~$X^*$ to~${F((\epsilon))}^n$, which is the set defined
by the same formula that defines~$X$ (clearly which one is chosen is
immaterial).
Let $X$ denote our max-closed set, which is not necessarily closed.
The extension~$X^*$ of~$X$ is closed if and only if $X$ is,
however we will modify $X^*$ slightly
to obtain a closed semilinear max-closed set $\tilde X\subset
{F((\epsilon))}^n$ such that $\tilde X \cap F^n = X$. Our intention is to apply~(2)
to~$\tilde X$, and write it as an intersection
of basic max-closed sets. This and the relation~$\tilde X \cap F^n = X$ are not prima
facie sufficient to recover a similar expression for~$X$, because the
coefficients appearing in the definition of~$\tilde X$ are, in general,
elements of~$F((\epsilon))$. Nevertheless we will be able to
reduce to the case of coefficients in~$F$ by a sequence of formal
manipulations.

The first step is the definition of~$\tilde X$. The only properties that
we require are that $\tilde X$ must be closed max-closed semilinear
and that~$\tilde X \cap F^n = X$. To this aim, write $X$ as a finite union
\[
X = X_1 \cup \dotsb \cup X_m
\]
of relatively open semilinear sets. Call $H_i$ the affine subspace
generated by~$X_i$ for~$i\in\{1\dotsc m\}$, so by definition $X_i$ is open
in~$H_i$. Now define
\[
\tilde X_i = \overline{\big\{p\in H^*_i \;{\big|}\; \operatorname{B}_\epsilon(p) \cap H^*_i \subset
X^*_i\big\}}
\]
where $\operatorname{B}_\epsilon(p)$ denotes the ball of radius~$\epsilon$ centered
at~$p$ in the sup~distance, and the overline denotes the topological
closure. Clearly $X_i \subset \tilde X_i \subset X^*_i$, where the first
inclusion follows from $X_i$ being relatively open. Hence the following
set
\[
\tilde X = \operatorname{max-closure}\left(\tilde X_1 \cup \dotsb \cup
\tilde X_k \right)
\]
meets our requirements (where $\operatorname{max-closure}$ denotes
the closure under the maximum operation).

By (2), we can write $\tilde X$ as an intersection of basic closed
max-closed subsets of~${F((\epsilon))}^n$. So it suffices to prove that given a
basic closed max-closed~$B\subset {F((\epsilon))}^n$, then $B'=B\cap F^n$ is an
intersection of basic max-closed subsets of~$F^n$. Let $B'$ be
\[
   B' = \big\{ (x_1\dotsc x_n)\in F^n \;{\big|}\; a_1 x_1 + \dotsb + a_n x_n \ge c \big\}
\]
with $a_1\dotsc a_n,c\in F((\epsilon))$ satisfying the condition $a_i\ge 0$
for~$i\neq k_B$. We will prove that $B'$ is a finite positive Boolean
combination of sets of the form
\[
  \big\{ (x_1\dotsc x_n)\in F^n \;{\big|}\; a'_1 x_1 + \dotsb + a'_n x_n \succ c' \big\}
\]
with $a'_1\dotsc a'_n,c'\in F$ all satisfying $a'_i\ge 0$ for~$i\neq k_B$.
From this follows our assertion.

Multiplying the equation
\[
	a_1 x_1 + \dotsb + a_n x_n \ge c
\]
by a suitable power of~$\epsilon$ we can assume that all the
coefficients~$a_i$ are in~$F[[\epsilon]]$.
We will proceed by induction on the number of the coefficients $a_i$ which
are not in~$F$. If this number is~$0$, the only problem is that $c$ may
not be in~$F$. Either $c$ has infinite magnitude (i.e., 
the leading monomial has negative degree), or it is of the form
$c=c_0+\epsilon c_{>0}$ with~$c_0\in F$ and~$c_{>0}\in F[[\epsilon]]$. In
the first case, $B'$ can only be all of~$F^n$ or the empty set. In the
second case, $B'$ is
\[
B' = \big\{ (x_1\dotsc x_n)\in F^n \;{\big|}\; a_1 x_1 + \dotsb + a_n x_n \succ c_0 \big\}
\]
where $\succ$ is~$\ge$ if~$c_{>0}\le 0$, and $\succ$ is~$>$ if~$c_{>0}>0$.

Assume that the cardinality of the set $I\eqdef\{i\;{|}\; a_i\notin
F\}$ is a positive number~$m$. Write $a_i = (a_i)_0 + \epsilon (a_i)_{>0}$ with
$(a_i)_0\in F$. As before, if $c$ has infinite magnitude, then $B'$ is
trivial, hence we can also write~$c=c_0+\epsilon c_{>0}$.
Then $x\in F^n$ is an element of~$B'$ if and only if
\begin{align*}
	(a_1)_0 x_1 + \dotsb &+ (a_n)_0 x_n \ge c_0 \\
	&{\wedge} \\
	\big( \;\; (a_1)_0 x_1 + \dotsb + (a_n)_0 x_n > c_0
	\quad &{\vee} \quad
	(a_1)_{>0} x_1 + \dotsb + (a_n)_{>0} x_n \ge c_{>0}
	\;\; \big)
\end{align*}
The first and the second conditions in this expression are immediately of
the required form. The third condition is relevant only when
\[
	(a_1)_0 x_1 + \dotsb + (a_n)_0 x_n = c_0
\]
because if~$<$ holds then~$x\notin B'$ by the first condition, and if~$>$
holds then~$x\in B'$ by the second condition. We will bring this condition
in an equivalent form with less than~$m$ coefficients which are not
in~$F$.

Let $h$ be such that~$\abs{(a_h)_{>0}}$ is maximal, clearly~$h\in I$.
Dividing
\[
(a_1)_{>0} x_1 + \dotsb + (a_n)_{>0} x_n \ge c_{>0}
\]
by~$\abs{(a_h)_{>0}}$ we obtain an equivalent condition
such that the coefficient of~$x_h$ is~$\pm1\in F$. However, in order to
apply the inductive hypothesis, we also need to ensure that all the
coefficients except that of~$x_k$ are non negative. This can be achieved
by adding a suitable multiple of
\[
	(a_1)_0 x_1 + \dotsb + (a_n)_0 x_n = c_0
\]
to obtain
\begin{multline*}
	\left( \frac{(a_1)_{>0}}{\abs{(a_h)_{>0}}} + N\cdot(a_1)_0 \right) x_1 + \dotsb +
	\left( \frac{(a_n)_{>0}}{\abs{(a_h)_{>0}}} + N\cdot(a_n)_0 \right) x_n
	\ge\\
	\frac{c_{>0}}{\abs{(a_h)_{>0}}} + N \cdot c_0
\end{multline*}
in fact, $(a_i)_{>0}$ can be negative only if $(a_i)_0$ is positive.
\end{proof}

\begin{proof}[Proof of Theorem~\ref{th-normal-form} (3)--(4)]
Clearly the relations in our finite basis are semilinear, closed,
and max-closed, hence the {\it only if\/} of~(3) is immediate.
Topological closure
is preserved by finite intersections. Therefore, to establish the {\it
only-if\/}~part for~(4), we only have to check that projections of
semilinear closed sets are closed. For a contradiction, pick a point~$x_0$ in
the boundary of~$\pi(X)$, then for all~$\epsilon>0$ there is a
$x(\epsilon)\in X$ such that $\abs{\pi(x(\epsilon))-x}_\infty<\epsilon$.
As usual, by definable choice, we can assume that the
function~$\epsilon\mapsto x(\epsilon)$ is semilinear, and taking the
limit for~$\epsilon\to0$ we get an~$x\in X$ such that~$\pi(x)=x_0$.

For the {\it if\/}~part, first we prove~(4). By~(2), it suffices to show
that basic closed max-closed sets are primitive positive definable using our relations. Consider the set
\[
  X = \bigcup_i
    \big \{ (x_1\dotsc x_n) \;{\big|}\; a_{i,1}x_1 + \dotsb + a_{i,n}x_n \ge c_i \big \}
\]
with $a_{i,j}\ge 0$ for all~$i$ and all~$j\neq k$. We can write
equivalently $X=\bigcup_{i=1}^m X_i$ where $X_i$ denotes the set defined
by the formula
\[
\beta_ix_k \le \sum_j \alpha_{i,j}x_j - c_i
\]
and the coefficients $\beta_i$, $\alpha_{i,j}$, and~$c_i$ are chosen in
such a way that they are all integers with $\beta_i>0$
and~$\alpha_{i,j}\ge0$.

First we prove that $X_i$ is primitive positive definable using~$\{1,-1,S_1,S_2\}$. To this aim, it suffices to show
primitive positive definitions of the sets defined by $2^Mx\le y$ and~$y\le
z_1+\dotsb+z_N$ for any $M$ and~$N$.
In fact, intersecting them and projecting along~$y$ one obtains the
set~$2^Mx\le z_1+\dotsb+z_N$, and assigning $x_1\dotsc x_n$ or~$\pm1$ to
the variables $x$ and~$z_1\dotsc z_N$ it is easy to obtain~$X_i$. The
set~$2^Mx\le y$ is defined inductively on~$M$ by
\[
{2^{M+1}\le y}\quad{\Leftrightarrow}\quad{\exists t \; 2x\le t \wedge 2^Mt\le y}
\]
And the set $y\le z_1+\dotsb+z_N$ is defined inductively on~$N$ by
\[
{y\le z_1+\dotsb+z_{N+1}}\quad{\Leftrightarrow}\quad
{\exists t \; y\le t+z_{N+1} \wedge t\le z_1+\dotsb+z_N}
\]

Now we show that the union of the sets~$X_i$ is primitive positive definable. First we
replace the variable $x_k$ on the left hand side in the definition of each
of these sets (remember that $k$ is fixed) with a new variable~$x'_i$.
Thus we end up with the following sets
\[
{(x'_i,x_1\dotsc x_n)\in X'_i}\quad{\Leftrightarrow}\quad
{\beta_ix'_i \le \sum_j \alpha_{i,j}x_j - c_i}
\]
remember that we assumed, without loss of generality, that all
coefficients~$\beta_i$ are strictly positive. We can combine the
sets~$X'_i$ to form a new definition of~$X$
\[
{(x_1\dotsc x_n)\in X}\quad{\Leftrightarrow}\quad
\exists x'_1\dotsc x'_m \;
\begin{cases}
(x'_1,x_1\dotsc x_n)\in X'_1\\
\quad\quad\quad\wedge\\
(x'_m,x_1\dotsc x_n)\in X'_m\\
\quad\quad\quad\wedge\\
x_k \le x'_1 \vee \dotsb \vee x_k \le x'_m
\end{cases}
\]
Now it remains to show a primitive positive definition of the last clause
\[
x_k \le x'_1 \vee \dotsb \vee x_k \le x'_m
\]
We can proceed by induction on~$m$ observing that it is equivalent to
\[
\exists t \; (x_k\le t \vee x_k\le x'_m) \wedge
(t \le x'_1 \vee \dotsb \vee t \le x'_{m-1})
\]

The {\it if\/}~part of~(3) is analogous, using~(1) in place of~(2), and observing that the
relations~$x<y\vee x\le z$ and~$x<y \vee x<z$ are primitive positive definable.
\end{proof}


\begin{proof}[Proof of Theorem~\ref{th-syntax-tropical}]
We will concentrate on point~(1). Point~(2) is analogous to
Theorem~\ref{th-normal-form}~(3)--(4).
Denote by~$\tau_k\colon\Q^n\to\Q^n$ the translation by~$k\in\Q$ applied
componentwise
\[
\tau_k\colon (x_1\dotsc x_n) \mapsto (x_1+k\dotsc x_n+k)
\]
Applying Theorem~\ref{th-normal-form}, we can write $X$ as an intersection
of basic max-closed sets $X_i$. Let $X'_i$ denote the set
\[
	X'_i = \bigcap_{k\in\Q} \tau_k(X_i)
\]
Since $X$ is translation invariant, for all~$i$, we have $X\subset X'_i
\subset X_i$, hence $X$ is the intersection of the sets~$X'_i$. The
sets~$X'_i$ are clearly translation invariant, therefore we only need
to prove that each of them is an intersection of basic tropically convex
sets. In other words, given a basic max-closed set~$B$, we want to prove
that
\[
B' \eqdef \bigcap_{k\in\Q} \tau_k(B)
\]
is a finite intersection of basic tropically convex sets.
Let $B$ be
\[
  B = \bigcup_j
    \big \{ (x_1\dotsc x_n) \;{\big|}\; a_{j,1}x_1 + \dotsb + a_{j,n}x_n \succ_j c_j \big \}
\]
with $a_{j,l}\ge 0$ for all~$l\neq k_B$. It suffices to write $B'$ as a
positive Boolean combination of sets of the form
\[
   \big \{ (x_1\dotsc x_n) \;{\big|}\; a'_{j,1}x_1 + \dotsb + a'_{j,n}x_n \succ'_j c'_j \big\}
\]
with $\sum_l a'_{j,l}=0$ and $a'_{j,l}\ge 0$ for all~$l\neq k_B$. By definition we have
\[
  B' = \bigcap_{k\in\Q} \bigcup_j
    \big \{ (x_1\dotsc x_n) \;{\big|}\; a_{j,1}x_1 + \dotsb + a_{j,n}x_n - k s_j \succ_j c_j \big \}
\]
where $s_j = \sum_l a_{j,l}$. We can separate the indices~$j$
satisfying~$s_j=0$ writing $B'=B'_0\cup B'_{\neq0}$ where
\begin{align*}
B'_0 &= \bigcup_{j|s_j=0}
\big\{ (x_1\dotsc x_n) \;{\big|}\; a_{j,1}x_1 + \dotsb + a_{j,n}x_n \succ_j c_j \big\} \\
B'_{\neq0} &= \bigcap_{k\in\Q} \bigcup_{j|s_j\neq 0}
\big\{ (x_1\dotsc x_n) \;{\big|}\; a_{j,1}x_1 + \dotsb + a_{j,n}x_n - k s_j \succ_j c_j \big\}
\end{align*}
Now, $B'_0$ is already in the required form.
Rearranging the definition of~$B'_{\neq0}$ we get
\[
x\in B'_{\neq0} \Leftrightarrow \Q = \bigcup_{j|s_j\neq0} \big\{ k\in\Q \;{\big|}\;
a_{j,1}x_1 + \dotsb + a_{j,n}x_n - c_j \succ_j k s_j \big\}
\]
On the right hand side, we have a union of left and right (according to
the sign of~$s_j$) half lines. This union covers all of~$\Q$ if and only
if it contains two opposite overlapping half lines.
Hence
\[
B'_{\neq0} = \bigcup_{j,l|s_j<0\wedge s_l>0}
\left\{ (x_1\dotsc x_n) \;{\Big|}\;
\sum_{i=1}^n (a_{j,i}s_l - a_{l,i}s_j) x_i \succ_{j,l} c_js_l - c_ls_j
\right\}
\]
where $\succ_{j,l}$ is~$>$ if $\succ_j$ and~$\succ_l$ are both $>$,
and it is~$\ge$ otherwise.
\end{proof}


\section{A Duality for Max-plus-average Inequalities}
\label{sect:duality}

\noindent Let $\mathcal{O}_n$ be the class of functions mapping
$\big(\Q\cup\{+\infty\}\big)^n$ to~$\Q\cup\{+\infty\}$ of either of the
following forms
\begin{align*}
(x_1\dotsc x_n) &\mapsto
	\max(x_{j_1}+k_1\dotsc x_{j_m}+k_m) \\
(x_1\dotsc x_n) &\mapsto
	\min(x_{j_1}+k_1\dotsc x_{j_m}+k_m) \\
(x_1\dotsc x_n) &\mapsto
	\frac{\alpha_1 x_{j_1} + \dotsb + \alpha_m
	x_{j_m}}{\alpha_1+\dotsb+\alpha_m} + k
\end{align*}
where~$k,k_i\in\Q$ and~$\alpha_i\in\Q^{>0}$.

For any given vector~$\bar o\in\mathcal{O}_n^n$ of $n$ operators in~$\mathcal{O}_n$, we consider
the following satisfiability problems: the {\it primal\/}~$P(\bar o)$
and the {\it dual\/}~$D(\bar o)$
\[
{
P(\bar o)\colon\begin{cases}
        \bar x \in \Q^n \\
        \bar x < \bar o (\bar x)
\end{cases}
}\quad\quad\quad
{
D(\bar o)\colon\begin{cases}
        \bar y \in \big(\Q \cup \{+\infty\}\big)^n \setminus \{+\infty\}^n \\
        \bar y \ge \bar o (\bar y)
\end{cases}
}
\]
where $<$ and~$\ge$ are meant to hold component-wise.
We intend to prove the following result.

\begin{theorem}\label{th-duality}
For any~$\bar o\in\mathcal{O}_n^n$ one and only one of the
problems~$P(\bar o)$ and~$D(\bar o)$ is satisfiable.
\end{theorem}

For the proof of Theorem~\ref{th-duality},
we will make use of zero-sum stochastic games with perfect information, in
the flavours known as the discounted and the limiting average payoff. In
this context, it is more natural to work over the domain~$\R$ of the real
numbers instead of~$\Q$. Observing that the satisfiability of~$P(\bar o)$
and~$D(\bar o)$ is insensitive to the domain, for the rest of this
section, we will work with the real numbers.

The set-up for a stochastic game is a directed graph~$G$ with
vertex-~and~edge-labels, all vertices of~$G$ must have at least one
out-edge. Each vertex of~$G$ is either assigned to one of the players, in
this case it is labelled with one of the symbols~\pmax\ and~\pmin, or it is
a {\it stochastic vertex}, and in this case it carries the label~\vs. Each
edge~$e$ of~$G$ has a label~$\po(e)\in\R$, which represents the payoff
earned by~\pmax\ (or the penalty incurred by~\pmin) when that edge is
traversed. The out-edges of a stochastic vertex have an additional
label~$\pr(e)$, which is a rational number representing the probability
that each edge is taken when exiting that specific vertex: clearly, the
probabilities of the out-edges of each stochastic vertex must sum to~$1$.

A stochastic game~$G$ is played by moving a token along the directed edges of~$G$.
First the token is placed on a vertex which we call the~{\it initial
position}. Then, repeatedly, the token is moved by~\pmax\ when it is on a
\pmax-vertex, by~\pmin\ when it is on a \pmin-vertex, and at random when it
is on an \vs-vertex (according to the probabilities
assigned to the out-edges of that vertex). A play never ends.

\def\v{\mathbf{v}}
Let $e_1,e_2\dotsc$ be the edges traversed during a play~$p$ on~$G$.
The {\it discounted payoff}~$\v_\beta(p)$ of~$p$ with discounting
factor~$\beta\in[0,1[$ is
\[
\v_\beta(p) \eqdef (1-\beta) \sum_{i=1}^{\infty} \po(e_i)\beta^{i-1}
\]
and the {\it limiting average payoff} is
\[
\v_1(p) \eqdef \liminf_{T\to\infty} \frac{1}{T}\sum_{i=1}^T \po(e_i)
\]
In such formulas we suppress the dependency on~$G$, to
ease the notation.

In general, a strategy for player~$P\in\{\pmax,\pmin\}$ is a probability
distribution on the functions
mapping every partial play~$v_0\overset{e_1}{\rightarrow} v_1 \dotsc
\overset{e_T}{\rightarrow} v_T$ ending in a
$P$-vertex~$v_T$ to one of the out-edges of~$v_T$: such functions are
called {\it pure strategies}.
Playing according to a strategy~$\pi$ means to choose one of the pure
strategies~$f$ at random with the probability distribution~$\pi$, and
then
choosing one's moves by calling~$f$. Given two
strategies~$\pi_\pmax$ and~$\pi_\pmin$ for the two players, one can thus
define the value~$\v_\beta(v,\pi_\pmax,\pi_\pmin)$ as the
expected~$\v_\beta$ of a play generated by the given pair of strategies
from the initial position~$v_0=v$.

For a given game~$G$ and $\beta\in[0,1]$, a strategy~$\pi_\pmax$
for~\pmax\ is
optimal if for all vertices~$v$ it
maximizes the quantity
\[
\v_\beta(v,\pi_\pmax,\cdot) \eqdef
\inf_{\pi_\pmin}\v_\beta(v,\pi_\pmax,\pi_\pmin)
\]
Conversely a strategy~$\pi_\pmin$ for~$\pmin$ is optimal if it minimizes
\[
\v_\beta(v,\cdot,\pi_\pmin) \eqdef
\sup_{\pi_\pmax}\v_\beta(v,\pi_\pmax,\pi_\pmin)
\]
Given a
stochastic game~$G$ and a $\beta\in[0,1]$,
there are optimal strategies for each player, moreover, calling
$\pi_\pmax$ and~$\pi_\pmin$ two optimal strategies
\[
\v_\beta(v,\pi_\pmax,\cdot) = \v_\beta(v,\cdot,\pi_\pmin)
\]
holds for any initial position~$v$. This value does not depend on the
pair of optimal strategies chosen, but only on~$\beta$ and the initial
position~$v$ (and~$G$), hence we may denote it by the
notation~$\v_\beta(v)$.
The
vector~$\v_\beta\eqdef(\v_\beta(v))_{v\in\vertices(G)}$ is called
{\it value vector}
of~$G$ with discount factor~$\beta$.

In the
specific case of perfect information stochastic games, one can show that
there are optimal strategies which are also pure and {\it stationary}
(also known as {\it memoryless} or {\it positional\/}).
In other words there are optimal strategies that prescribe for each
possible position of the token (i.e.\ each \pmax- or \pmin-vertex) one definite
next move. If we further restrict our consideration to the discounted
payoff criterion (i.e.~$\beta<1$), then the value vector of a game~$G$
admits an explicit description through a condition known as the {\it limit discount
equation} (see~\cite{Filar-Vrieze} Theorem~4.3.2 and also Definition~4.3.12)
\[
\v_\beta(v) = \begin{cases}
\max_{(v,w)\in\edges(G)} (1-\beta) \po(v,w) + \beta \v_\beta(w)
& \text{\phantom{if $v$ is a \pmin-vertex}\hbox to 0pt{\hss if $v$ is a \pmax-vertex}} \\
\min_{(v,w)\in\edges(G)} (1-\beta) \po(v,w) + \beta \v_\beta(w) &\text{if
$v$ is a \pmin-vertex} \\
\mathrlap{
\sum_{(v,w)\in\edges(G)} \pr(v,w) \big((1-\beta) \po(v,w) + \beta
\v_\beta(w)\big)} \\
& \text{\phantom{if $v$ is a \pmin-vertex}\hbox to 0pt{\hss if $v$ is a stochastic vertex}}
\end{cases}
\]
In particular we will need the following fact (see~\cite{Filar-Vrieze}
Theorem~6.3.7 plus Theorem~6.3.5 and its proof).
\begin{fact}\label{th-fact-games}For any stochastic game~$G$ with perfect information
\begin{enumerate}
\item\label{point-stationary}both players possess
pure stationary strategies~$\pi_\pmax$ and~$\pi_\pmin$ which are optimal
for the limiting average payoff and for
all discount factors~$\beta$ sufficiently close to~$1$
\item\label{point-limit}calling $\v_\beta$ the value vector of~$G$ with discount factor~$\beta$,
the value vector for the limiting average payoff can be
written as $\v_1 = \lim_{\beta\to1}\v_\beta$
\item\label{point-series}the solution~$\v_\beta$ to the limit discount equation can be
written as a power series in~$(1-\beta)$---more precisely, consider the
field $\R((x))$ of formal Laurent series in~$x$ ordered by~$0<x\ll1$, let
$\beta=1-x\in\R((x))$, then the limit discount equation for~$\v_\beta$ has a formal
solution in~$\R((x))^n$ which, moreover, is convergent in a neighbourhood
of~$0$.
\end{enumerate}
\end{fact}

An alternative approach to ours
would have been to use the normal form described in~\cite{BEGM}. In fact,
probably, the duality and the existence of a normal form in the sense
of~\cite{BEGM} imply each other. However we obtain our result via a
different method.

We now map each vector of operators~$\bar o \in\mathcal{O}_n^n$
to a stochastic game~$G_{\bar o}$. First we place one vertex in~$G_{\bar
o}$ per component of~$\bar o$, i.e.\ formally we fix~$\vertices(G_{\bar o}) =
\{v_1\dotsc v_n\}$. Then we stipulate that $v_i$ is of type $\pmax$,
$\pmin$, or~$\vs$ according to whether $\bar o_i$ is a $\max$, $\min$,
or weighted average operator respectively. Finally, for each~$i$, let
\[
\bar o_i \colon (x_1\dotsc x_n) \mapsto \square_i (x_{j^i_1}+k^i_1\dotsc
x_{j^i_{m^i}}+k^i_{m^i})
\]
where $\square_i$ can be either of
\[
\square_i \colon (y_1\dotsc y_{m^i}) \mapsto
\begin{cases}
\max(y_1\dotsc y_{m^i}) \\
\min(y_1\dotsc y_{m^i}) \\
\frac{y_1\alpha^i_1 + \dotsb +
y_{m^i}\alpha^i_{m^i}}{\alpha^i_1+\dotsb+\alpha^i_{m^i}}
\end{cases}
\]
then we introduce an edge~$(v_i,v_{j^i_l})\in\edges(G_{\bar o})$ for each pair of~$i\in\{1\dotsc n\}$
and~$l\in\{1\dotsc m_i\}$ with payoff
\[\po(v_i,v_{j^i_l})=k^i_l\]
and, if $\square_i$
is a weighted average, with probability
\[\pr(v_i,v_{j^i_l}) =
\frac{\alpha^i_l}{\alpha^i_1+\dotsb+\alpha^i_{m^i}}
\]

\begin{lemma}\label{th-primal-game}
Let $\bar o\in \mathcal O_n^n$ be a vector of operators, and let $\v_1$
denote the value vector of the game~$G_{\bar o}$ with the limiting average
payoff.
The problem~$P(\bar o)$ is satisfiable if and only if
$\v_1(v_i)>0$ for all vertices~$v_i$ of~$G_{\bar o}$.
\end{lemma}
\begin{proof}[Proof (if direction)]
Using Fact~\ref{th-fact-games}(\ref{point-series}), let
\[
	\v_\beta = \sum_{i=0}^\infty \bar a_i(1-\beta)^i
\]
be the value vector of~$G_{\bar o}$ with discount~$\beta$. It follows from
Fact~\ref{th-fact-games}(\ref{point-limit}) that $(\bar a_0)_j =
\v_1(v_j)>0$ for all~$j\in\{1\dotsc n\}$.
For~$N$ denoting a (large) real number, define $\bar x_N\in\R^n$ by
\[
	(\bar x_N)_j = N (\bar a_0)_j + (\bar a_1)_j
\]
We claim that, for large enough~$N$, the vector~$\bar x_N$ satisfies~$P(\bar o)$.

Let us consider the limit discount equation for \pmax-, \pmin-,
and~\vs-vertices separately.
We will show that the condition on each vertex~$v_j$ implies that the
corresponding coordinate~$(\bar x_N)_j$ satisfies~$P(\bar
o)$.

If $v_j$ is of type \pmax, then
\[
\sum_{i=0}^\infty (\bar a_i)_j(1-\beta)^i =
\max_{l\;|\;(v_j,v_l)\in\edges(G_{\bar o})} \big(
(1-\beta) \po(v_j,v_l) + \beta
\sum_{i=0}^\infty (\bar a_i)_l(1-\beta)^i
\big)
\]
Since series are ordered lexicographically, truncating both sides to the
first two terms commutes with the $\max$ operation
\begin{multline*}
(\bar a_0)_j +
(\bar a_1)_j(1-\beta) =\\
\max_{l\;|\;(v_j,v_l)\in\edges(G_{\bar o})}
(\bar a_0)_l
+
\big(
\po(v_j,v_l) - (\bar a_0)_l + (\bar a_1)_l
\big)
(1-\beta)
\end{multline*}
From which it follows that
\begin{align*}
(\bar a_0)_j & = \max_{l\;|\;(v_j,v_l)\in\edges(G_{\bar o})} (\bar a_0)_l \\
(\bar a_1)_j & =
\max_{\text{$l$ realizing the $\max$ above}}
\po(v_j,v_l) - (\bar a_0)_l + (\bar a_1)_l \\
& <
\max_{\text{$l$ realizing the $\max$ above}}
\po(v_j,v_l) + (\bar a_1)_l
\end{align*}
Hence, for large values of~$N$
\[
N (\bar a_0)_j + (\bar a_1)_j <
\max_{l\;|\;(v_j,v_l)\in\edges(G_{\bar o})}
N (\bar a_0)_l + \po(v_j,v_l) + (\bar a_1)_l
\]

Vertices of type \pmin\ are treated precisely as those of type~\pmax.
Finally, for stochastic vertices, we get
\begin{multline*}
(\bar a_0)_j +
(\bar a_1)_j(1-\beta) =\\
\sum_{l\;|\;(v_j,v_l)\in\edges(G_{\bar o})}
(\bar a_0)_l
+
\big(
\po(v_j,v_l) - (\bar a_0)_l + (\bar a_1)_l
\big)
(1-\beta)
\end{multline*}
whence, multiplying by~$(1-\beta)^{-1}$ and evaluating at $1-\beta = 1/N$
\begin{align*}
N (\bar a_0)_j +
(\bar a_1)_j & =
\sum_{l\;|\;(v_j,v_l)\in\edges(G_{\bar o})}
N (\bar a_0)_l
+
\po(v_j,v_l) - (\bar a_0)_l + (\bar a_1)_l \\
&<
\sum_{l\;|\;(v_j,v_l)\in\edges(G_{\bar o})}
N (\bar a_0)_l
+
\po(v_j,v_l) + (\bar a_1)_l
\end{align*}
\end{proof}
\begin{proof}[Proof (only if direction)]
Fix a solution~$\bar x$ of~$P(\bar o)$.
We claim that the following strategy~$\pi_\pmax$ for~\pmax\ (which,
incidentally, is stationary and pure) satisfies
$\v_1(v_i,\pi_\pmax,\cdot)>0$ for all vertices~$v_i$ of~$G_{\bar o}$.
Let~$v_i$ be a \pmax-vertex, assume
that it is \pmax's turn and the token rests on~$v_i$, we
will say which of the out-edges of~$v_i$ the player~\pmax\ will elect to move the
token along.
Let us consider the operator~$\bar o_i$, which
necessarily has the following form
\[
\bar o_i \colon (x_1\dotsc x_n) \mapsto \max (x_{j^i_1}+k^i_1\dotsc
x_{j^i_{m^i}}+k^i_{m^i})
\]
The player \pmax\ then moves the token to a vertex~$v_{j^i_l}$ such that
$\bar x_{j^i_l}+k^i_l$ realizes the maximum (which one he chooses is
immaterial).

Now we intend to prove that $\v_1(v_i,\pi_\pmax,\cdot)> 0$.
To this aim, because of
Fact~\ref{th-fact-games}(\ref{point-stationary}), it is enough to test our
strategy~$\pi_\pmax$ against all stationary and pure strategies for~\pmin.
Let $\pi_\pmin$ be such a strategy.
A play generated by the pair of strategies~$(\pi_\pmax,\pi_\pmin)$ is a
Markov chain on the finite state-space~$\{v_1\dotsc v_n\}$.
The average time spent by the process in each state during plays starting
at~$v_i$ must converge to a stable distribution~$\mu$.
The limiting average payoff~$\v_1(v_i,\pi_\pmax,\pi_\pmin)$ can
be described using~$\mu$ by
\[
\v_1(v_i,\pi_\pmax,\pi_\pmin) = \sum_{a,b\;|\;(v_a,v_b)\in\edges(G_{\bar o})}
\mu(v_a)\pr'(v_a,v_b)\po(v_a,v_b)
\]
where
\[
\pr'(v_a,v_b) = \begin{cases}
\pr(v_a,v_b) &\text{if $v_a$ is a stochastic vertex}\\
1	&\text{if $v_a$ is a \pmax-vertex and $\pi_\pmax(\dotsb\to v_a) = v_b$}\\
1	&\text{if $v_a$ is a \pmin-vertex and $\pi_\pmin(\dotsb\to v_a) = v_b$}\\
0	&\text{otherwise}
\end{cases}
\]
We need to prove that the above quantity is strictly positive.

For all $a\in\{1\dotsc n\}$ we have
\[
\bar x_a < \sum_{b\;|\;(v_a,v_b)\in\edges(G_{\bar o})}
\pr'(v_a,v_b)(\bar x_b + \po(v_a,v_b))
\]
which, for \pmin- and \vs-vertices, is an immediate consequence of $\bar x$
being a solution to~$P(\bar o)$, and, when $v_a$ is a \pmax-vertex, it
follows for our choice of the strategy~$\pi_\pmax$. Splitting the sum
on the right hand side
\[
\bar x_a - \sum_{b\;|\;(v_a,v_b)\in\edges(G_{\bar o})}
\pr'(v_a,v_b)\bar x_b
<
\sum_{b\;|\;(v_a,v_b)\in\edges(G_{\bar o})}
\pr'(v_a,v_b)\po(v_a,v_b)
\]
Then multiplying by $\mu(v_a)$ and taking the sum over~$a$
\begin{multline*}
\sum_{a=1}^n \mu(v_a)\bar x_a
- \sum_{a,b\;|\;(v_a,v_b)\in\edges(G_{\bar o})}
  \mu(v_a)\pr'(v_a,v_b)\bar x_b\\
<\sum_{a,b\;|\;(v_a,v_b)\in\edges(G_{\bar o})}
\mu(v_a)\pr'(v_a,v_b)\po(v_a,v_b)
\end{multline*}
We need to prove that the left hand side is not negative.
By $\mu$ being a stable distribution, we have
\[
\mu(v_b) = \sum_{a\;|\;(v_a,v_b)\in\edges(G_{\bar o})} \mu(v_a)\pr'(v_a,v_b)
\]
hence
\begin{align*}
\sum_{a=1}^n \mu(v_a)\bar x_a
- &\sum_{a,b\;|\;(v_a,v_b)\in\edges(G_{\bar o})}
  \mu(v_a)\pr'(v_a,v_b)\bar x_b \\
&= \sum_{a=1}^n \mu(v_a)\bar x_a
- \sum_{b=1}^n \big( \sum_{a\;|\;(v_a,v_b)\in\edges(G_{\bar o})}
\mu(v_a)\pr'(v_a,v_b) \big)  \bar x_b\\
&= \sum_{a=1}^n \mu(v_a)\bar x_a
- \sum_{b=1}^n \mu(v_b)\bar x_b = 0
\end{align*}
\end{proof}

\begin{lemma}\label{th-dual-game}
Let $\bar o\in \mathcal O_n^n$ be a vector of operators, and let $\v_1$
denote the value vector of the game~$G_{\bar o}$ with the limiting average
payoff.
The problem~$D(\bar o)$ is satisfiable if and only if
$\v_1(v_i)\le0$ for some vertex~$v_i$ of~$G_{\bar o}$.
\end{lemma}
The proof of this lemma is similar to the proof of
Lemma~\ref{th-primal-game} above, we will therefore only point out the
relevant differences.
\begin{proof}[Proof (if direction)]
As in the proof of Lemma~\ref{th-primal-game},
using Fact~\ref{th-fact-games}(\ref{point-series}), let
\[
	\v_\beta = \sum_{i=0}^\infty \bar a_i(1-\beta)^i
\]
be the value vector of~$G_{\bar o}$ with discount~$\beta$.
Let $\v_1^{\min}\le0$ denote the minimum
of~$\v_1(v_j) = (\bar a_0)_j$ over all vertices~$v_j$.
We define $\bar y\in(\R\cup\{+\infty\})^n$
satisfying~$D(\bar o)$
by
\[
	\bar y_j = \begin{cases}
		(\bar a_1)_j &\text{if $(\bar a_0)_j = \v_1^{\min}$} \\
		+\infty &\text{otherwise}
	\end{cases}
\]

We want to show that the condition imposed by the limit discount equation
on each vertex~$v_j$ implies that the
corresponding coordinate~$\bar y_j$ satisfies~$D(\bar
o)$. As opposed to Lemma~\ref{th-primal-game}, we can concentrate on those vertices that
have minimal value, i.e.~$(\bar a_0)_j = \v_1^{\min}$, because those with
larger value give rise to
$+\infty$ coordinates for which~$D(\bar o)$ is automatically satisfied.
If $v_j$ is of type \pmin, then, considering the limit discount equation
and
truncating the series to the first two terms, we have
\begin{multline*}
\v_1^{\min} +
(\bar a_1)_j(1-\beta) =\\
\min_{l\;|\;(v_j,v_l)\in\edges(G_{\bar o})}
(\bar a_0)_l
+
\big(
\po(v_j,v_l) 
- (\bar a_0)_l
+ (\bar a_1)_l
\big)
(1-\beta)
\end{multline*}
The $\min$ can be restricted to those values of~$l$ for which $(\bar
a_0)_l= \v_1^{\min}$
\begin{multline*}
\v_1^{\min} +
(\bar a_1)_j(1-\beta)
=\\
\min_{l\;|\;(v_j,v_l)\in\edges(G_{\bar o})\wedge(\bar a_0)_l= \v_1^{\min}}
\v_1^{\min}
+
\big(
\po(v_j,v_l) 
- \v_1^{\min}
+ (\bar a_1)_l
\big)
(1-\beta)
\end{multline*}
Subtracting $\v_1^{\min}$ from both sides and multiplying
by~$(1-\beta)^{-1}$
\begin{align*}
\bar y_j =
(\bar a_1)_j
&=
\min_{l\;|\;(v_j,v_l)\in\edges(G_{\bar o})\wedge(\bar a_0)_l= \v_1^{\min}}
\po(v_j,v_l) 
- \v_1^{\min}
+ (\bar a_1)_l \\
&\ge
\min_{l\;|\;(v_j,v_l)\in\edges(G_{\bar o})\wedge(\bar a_0)_l= \v_1^{\min}}
\po(v_j,v_l) 
+ (\bar a_1)_l \\
&=
\min_{l\;|\;(v_j,v_l)\in\edges(G_{\bar o})}
\po(v_j,v_l) 
+ \bar y_l
\end{align*}

For $v_j$ of type \pmax\ we obtain as before
\begin{multline*}
\v_1^{\min} +
(\bar a_1)_j(1-\beta) =\\
\max_{l\;|\;(v_j,v_l)\in\edges(G_{\bar o})}
(\bar a_0)_l
+
\big(
\po(v_j,v_l) 
- (\bar a_0)_l
+ (\bar a_1)_l
\big)
(1-\beta)
\end{multline*}
hence
\[
\v_1^{\min} = \max_{l\;|\;(v_j,v_l)\in\edges(G_{\bar o})} (\bar a_0)_l
\]
and, by the minimality of~$\v_1^{\min}$, we have that
$(\bar a_0)_l =
\v_1^{\min}$ for all~$l$ such that $(v_j,v_l)\in\edges(G_{\bar o})$.
Therefore
\begin{multline*}
\v_1^{\min} +
(\bar a_1)_j(1-\beta) =\\
\max_{l\;|\;(v_j,v_l)\in\edges(G_{\bar o})}
\v_1^{\min}
+
\big(
\po(v_j,v_l) 
- (\bar a_0)_l
+ (\bar a_1)_l
\big)
(1-\beta)
\end{multline*}
and we conclude again subtracting $\v_1^{\min}$ and multiplying
by~$(1-\beta)^{-1}$.

Finally stochastic vertices are dealt with as \pmax-vertices observing
that
\[
\v_1^{\min} = \sum_{l\;|\;(v_j,v_l)\in\edges(G_{\bar o})}
\pr(v_j,v_l)  (\bar a_0)_l
\]
implies, again by the minimality of~$\v_1^{\min}$, that 
$(\bar a_0)_l =
\v_1^{\min}$ for all~$l$ such that $(v_j,v_l)\in\edges(G_{\bar o})$.
\end{proof}
\begin{proof}[Proof (only if direction)]
Fix a solution~$\bar y\in
\big(\R \cup \{+\infty\}\big)^n \setminus \{+\infty\}^n$ of~$D(\bar o)$.
We produce a strategy~$\pi_\pmin$ for~\pmin\ that satisfies
$\v_1(v_i,\pi_\pmax,\cdot)\le 0$ for some~$v_i$: specifically
$\v_1(v_i,\pi_\pmax,\cdot)\le 0$ if $\bar y_i$ is finite.
Assume
that it is \pmin's turn and the token rests on~$v_i$. If $\bar y_i =
+\infty$, then it is immaterial which move \pmin\ chooses.
Otherwise, let us consider the operator
\[
\bar o_i \colon (x_1\dotsc x_n) \mapsto \min (x_{j^i_1}+k^i_1\dotsc
x_{j^i_{m^i}}+k^i_{m^i})
\]
The player \pmin\ moves to a vertex~$v_{j^i_l}$ such that
$\bar x_{j^i_l}+k^i_l$ realizes the $\min$.

As in the proof of Lemma~\ref{th-primal-game}, pick a stationary pure
strategy~$\pi_\pmax$ for~$\pmax$, and consider the Markov process
defined by~$\pi_\pmin$ and~$\pi_\pmax$ starting from vertex~$v_i$. If
$\bar y_i$ is finite, then it is easy to see that, by our choice
of~$\pi_\pmin$, no vertex~$\v_j$ with~$\bar y_j = +\infty$ can be reached
by a play starting from~$\v_i$. Let $\mu$ be the stable distribution to
which the Markov chain started from~$v_i$ converges.
The limiting average payoff is
\[
\v_1(v_i,\pi_\pmax,\pi_\pmin) = \sum_{a,b\;|\;(v_a,v_b)\in\edges(G_{\bar o})}
\mu(v_a)\pr'(v_a,v_b)\po(v_a,v_b)
\]
where
\[
\pr'(v_a,v_b) = \begin{cases}
\pr(v_a,v_b) &\text{if $v_a$ is a stochastic vertex}\\
1	&\text{if $v_a$ is a \pmax-vertex and $\pi_\pmax(\dotsb\to v_a) = v_b$}\\
1	&\text{if $v_a$ is a \pmin-vertex and $\pi_\pmin(\dotsb\to v_a) = v_b$}\\
0	&\text{otherwise}
\end{cases}
\]

For $a\in\{1\dotsc n\}$ such that $\bar y_a$ is finite, we have
\[
\bar y_a \ge \sum_{b\;|\;(v_a,v_b)\in\edges(G_{\bar o})}
\pr'(v_a,v_b)(\bar y_b + \po(v_a,v_b))
\]
and proceeding as in the proof of Lemma~\ref{th-dual-game}
\begin{multline*}
\sum_{a\;|\;\bar y_a \neq +\infty} \mu(v_a)\bar y_a
- \sum_{a,b\;|\;(v_a,v_b)\in\edges(G_{\bar o})\wedge \bar y_a \neq +\infty}
  \mu(v_a)\pr'(v_a,v_b)\bar y_b\\
\ge \sum_{a,b\;|\;(v_a,v_b)\in\edges(G_{\bar o})\wedge \bar y_a \neq +\infty}
\mu(v_a)\pr'(v_a,v_b)\po(v_a,v_b)
\end{multline*}
where the sums can be restricted to~$\bar y_a \neq +\infty$ because the
vertices~$v_a$ with $\bar y_a = +\infty$ are unreachable. We can conclude
using the fact that $\mu$ is a stable distribution.
\end{proof}

Theorem~\ref{th-duality} follows immediately from
Lemma~\ref{th-primal-game} and Lemma~\ref{th-dual-game}.

\section{Complexity of Tropically Convex CSPs}
\label{sect-tconv}
\label{sec-csp}
\def\NP{\mathsf{NP}}
\def\coNP{\text{co-}\NP}

\noindent In this section, we will apply our duality to tropically convex
constraint satisfaction problems. By Theorem~\ref{th-syntax-tropical}, we
know that the tropically convex relations are precisely
those primitive positive definable in the structure $\Gamma_t =
(\Q;<,T_1,T_{-1},S_3,M_0)$
where
\begin{align*}
T_{\pm1}(x,y) &\Leftrightarrow x \le y \pm 1 \\
S_3(x,y,z) &\Leftrightarrow x \le \frac{y+z}{2} \\
M_0(x,y,z) &\Leftrightarrow x \le \max(y,z)
\end{align*}
Max-atoms (with constants in binary) is polynomial-time reducible to
the CSP of~$\Gamma_t$, but $\Csp(\Gamma_t)$ is more expressive.
We intend to prove the following theorem.

\begin{theorem}\label{th-csp}
The problem $\Csp(\Gamma_t)$ is in $\NP\cap\coNP$.
\end{theorem}

Instead of the finite constraint language of~$\Gamma_t$, we could have chosen
to work with basic tropically convex sets (in the sense of
section~\ref{sect-syntax}) encoded with the constants expressed in binary.
In view of Observation~\ref{th-concise} this choice is immaterial. The
reader can also check that the same proofs apply to both settings.
Also, it is easy to extend $\Gamma_t$ with relations of the form~$x=c$ for rational
constants~$c$: the idea is to introduce a new variable~$z$ and replace
$x=c$ with~$x=c+z$ (which is primitive positive definable in~$\Gamma_t$), then
by translation invariance we have a solution if and only if we have a solution
with~$z=0$.
We
begin by proving an analogue of Theorem~\ref{th-duality} for non-strict
inequalities.

\begin{corollary}
\label{th-non-strict}
For any given vector of operators~$\bar o\in\mathcal{O}_n^n$ we consider
the following satisfiability problems: the {\it primal\/}~$P'(\bar o)$
and the {\it dual\/}~$D'(\bar o)$
\[
{
P'(\bar o)\colon\begin{cases}
        \bar x \in \Q^n \\
        \bar x \le \bar o (\bar x)
\end{cases}
}\quad\quad\quad
{
D'(\bar o)\colon\begin{cases}
        \bar y \in \big(\Q \cup \{+\infty\}\big)^n \setminus \{+\infty\}^n \\
        \bar y > \bar o (\bar y)
\end{cases}
}
\]
where $\le$ and~$>$ are meant to hold component-wise, and we stipulate
that~$+\infty>+\infty$.
Then one and only one of the
problems~$P'(\bar o)$ and~$D'(\bar o)$ is satisfiable.
\end{corollary}
\begin{proof}
We claim that the problem~$P'(\bar o)$ is satisfiable if and only if the
problem~$P_\epsilon(\bar o)$ defined as follows
\begin{align*}
	&\bar x \in \Q^n \\
	&\bar x < \bar o (\bar x) + \bar 1 \epsilon
\end{align*}
where $\bar 1$ denotes the vector~$(1,1\dotsc)\in\Q^n$, is satisfiable for
all~$\epsilon>0$. The only-if direction is, in fact, immediate. For the if
direction, by definable choice, we can choose a solution~$\bar x(\epsilon)$ of the
second problem which is a semilinear function of~$\epsilon$. Hence
the limit~$\bar x_0\eqdef\lim_{\epsilon\to0^+} \bar x(\epsilon)$ exists and it is easy to
check that $\bar x_0$ satisfies of~$P'(\bar o)$.

Applying Theorem~\ref{th-duality} we get that~$P_\epsilon(\bar o)$ is
satisfiable if and only if the following problem, $D_\epsilon(\bar o)$
\begin{align*}
	&\bar y \in \big(\Q \cup \{+\infty\}\big)^n \setminus \{+\infty\}^n \\
	&\bar y \ge \bar o (\bar y) + \bar 1 \epsilon
\end{align*}
is not satisfiable. In turn,
it is immediate that $D_\epsilon(\bar o)$ is satisfiable for some~$\epsilon>0$
if and only if $D'(\bar o)$ is satisfiable.
\end{proof}

Before proceeding to the proof of Theorem~\ref{th-csp} we need one last
technical statement.

\begin{definition}\label{def-zero-plus}
Consider a quantifier free
semilinear formula~$\phi(t,\bar x)$ with rational
coefficients, where $t$ denotes a variable and $\bar x$ denotes a tuple of variables.
We say that $\phi(t,\bar x)$ is {\it satisfiable in~$0+$} if
\[
\exists t_0 > 0 \; \forall t \in {]0,t_0]} \;\exists \bar x \;\phi(t,\bar x)
\]
\end{definition}

\begin{lemma}\label{th-germs}
The problem, given~$\phi$ as in Definition~\ref{def-zero-plus} with coefficients encoded in binary,
of deciding whether~$\phi$ is satisfiable in~$0+$ is in~$\NP$.
\end{lemma}
\begin{proof}
\def\beps{\boldsymbol\epsilon}
Satisfiability for quantifier free semilinear formulas is in~$\NP$
by standard linear programming techniques. In
our case, consider the two-dimensional ordered vector space over~$\Q$
\[
	V \eqdef \big\{a+b\beps \;{\big|}\; a,b\in\Q\big\}
\]
where the expression $a+b\beps$ is just a formal replacement for~$(a,b)$,
and the order is lexicographical -- in other words, writing $\beps$
for~$0+1\beps=(0,1)$, we have $(0,0)<(0,1)\ll(1,0)$ or~$0<\beps\ll1$. We claim that
$\phi(t,\bar x)$ is satisfiable in~$0+$ if and only if
$\phi(\beps,\bar x)$ is satisfiable in~$V$. Before proving the claim, we
may observe that this is, indeed, sufficient to conclude. In fact
\[
\exists \bar x \in V^n \; \phi(\beps,\bar x)
\]
is, by definition, equivalent to
\[
\exists \bar a, \bar b \in \Q^n \; \phi(\beps,\bar a + \bar b \beps)
\]
and the subformula~$\phi(\beps,\bar a + \bar b \beps)$ can be easily
replaced by a quantifier free semilinear formula over~$\Q$, by
substituting each basic relation with its component-wise definition.

It remains to prove the claim. At first, we can observe that  a basic
relation (say $a+b\beps < c+d\beps$) holds in~$V$, if and only if the corresponding
relation ($a+bt < c+dt$) holds for all~$t$
sufficiently small.
The if part of the claim follows immediately: let $\bar
x = \bar a+\bar b\beps$ be a satisfying assignment for~$\phi(\beps,\bar x)$ in~$V$,
then $\phi(t,\bar a+\bar bt)$ must hold for all sufficiently small
positive~$t\in\Q$. For the only if part,
if~$\phi(t,\bar x)$ is satisfiable for
all~$t\in{]0,t_0]}$, then, by the definability of Skolem
functions, there is a semilinear function
\[
	\bar f \colon ]0,t_0] \to \Q^n
\]
such that~$\bar x = \bar f(t)$ is a satisfying assignment for
all~$t\in{]0,t_0]}$. Now, for some
positive~$t_1<t_0$, the restriction of~$f$
to~$]0,t_1]$ must be linear. In other words, there are $\bar a$
and~$\bar b$ in~$\Q^n$ such that
\[
	\forall t\in{]0,t_1]} \; \phi(t,\bar a + \bar
b t)
\]
hence $\bar x = \bar a+\bar b\beps\in V$ must satisfy~$\phi(\beps,\bar x)$.
\end{proof}

\begin{proof}[Proof of Theorem~\ref{th-csp}]
For the $\NP$ part, it suffices to observe that after guessing
which of the inputs of each $\max$~constraint realizes the maximum, one is
left with a linear feasibility problem, which is in~$\mathsf P$.

For the $\coNP$ part, we will employ the following strategy. First we
replace each strict inequality~$A<B$ by~$\exists\epsilon>0\; A\le
B-\epsilon$. Hence we can apply Corollary~\ref{th-non-strict}, so we get
that our problem is not satisfiable if and only if for all~$\epsilon>0$
some~$D'(\bar o_\epsilon)$ is satisfiable. We then conclude by
Lemma~\ref{th-germs}.

First let $\phi$ denote an instance of~$\Csp(\Gamma_t)$.
Construct a new formula~$\phi_\epsilon$ by formally replacing each strict
inequality~$A<B$ in~$\phi$ by~$A\le B-\epsilon$, the new
formula~$\phi_\epsilon$ is hence a conjunction of non-strict
inequalities. Clearly $\phi$ is satisfiable if and only if there is
an~$\epsilon>0$ such that~$\phi_\epsilon$ is satisfiable. Now, the
formula~$\phi_\epsilon$ is almost in the form~$\bar x\le \bar o_\epsilon(\bar x)$
that we need to apply Corollary~\ref{th-non-strict}, except that the same
variable may appear in the left hand side of more that one constraint. To
correct this discrepancy, let $\Gamma'_t$ be the expansion of~$\Gamma_t$ obtained by addition of relations of the form
\[	x_0 \le \min(x_1\dotsc x_n) \]
We rewrite~$\phi_\epsilon$ as a new formula~$\phi'_\epsilon$ in the
language of~$\Gamma'_t$ by replacing each left hand side instance of a
variable~$x_i$
in~$\phi_\epsilon$ with a new variable~$x_{i,1},x_{i,2}\dotsc$ and then adding
a constraint
\[
	x_i \le \min(x_{i,1},x_{i,2}\dotsc)
\]
for each of the old variables~$x_i$. The formula~$\phi'_\epsilon$ is in
the form~$\bar x\le \bar o_\epsilon(\bar x)$, therefore, by
Corollary~\ref{th-non-strict} it is non-satisfiable if and only if $\bar y>
\bar o_\epsilon(\bar y)$ has a solution in~$\Q\cup\{+\infty\}$ which
is not~$+\infty$ on every coordinate. Hence we have reduced the
non-satisfiability of~$\phi$ to the following formula~$\psi$
\[
\forall\epsilon>0 \;\;\; \exists \bar y\in(\Q\cup\{+\infty\})^n\setminus\{+\infty\}^n
\;\;\; \bar y > \bar o_\epsilon(\bar y)
\]
which we intend to prove that can be checked in~$\NP$.
In fact, if an assignment $\bar y = \bar y_0$ satisfies
$\bar y > \bar o_\epsilon(\bar y)$ for~$\epsilon = \epsilon_0$, then the same
assignment must satisfy~$\bar y > \bar o_\epsilon(\bar y)$ also for any
other~$\epsilon>\epsilon_0$. As a consequence $\psi$ is equivalent to the
following formula~$\psi'$
\[
\exists \epsilon_0 > 0 \;\;\;
\forall\epsilon\in{]0,\epsilon_0]} \;\;\; \exists \bar y\in(\Q\cup\{+\infty\})^n\setminus\{+\infty\}^n
\;\;\; \bar y > \bar o_\epsilon(\bar y)
\]
We can therefore apply Lemma~\ref{th-germs}
observing that the domain~$\Q\cup\{+\infty\}$ can be coded
in~$(\Q;+,\le)$ in a quantifier free fashion---for instance by
pairs~$(a,b)\in\Q^2$ where $(x,1)$ represents the number~$x$ and $(1,0)$
represents~$+\infty$.
\end{proof}

\section{A Polynomial-time Tractable Fragment}
\label{sect:poly}
\noindent
We present an algorithm that tests satisfiability
of a given (quantifier-free) restricted Horn formula $\Phi$. 
Let $V$ be the set of variables of $\Phi$. 
Recall that each restricted Horn clause
has at most one literal which contains a
variable with a negative coefficient. 
We call this literal the \emph{positive literal}
of the clause, and all other literals the \emph{negative literals}. 

\vskip0.5\baselineskip
\begin{center}
\fbox{
\begin{tabular}{l}
Solve($\Phi$) \\
{\it // Input: a restricted Horn formula $\Phi$}\\
Do \\
\hspace{.5cm} Let $\Psi$ be the clauses in $\Phi$ that contain at most one literal.\\
\hspace{.5cm} If $\Psi$ is unsatisfiable then return {\it unsatisfiable}. \\
\hspace{.5cm}    For all negative literals $\varphi$ in clauses from $\Phi$ \\
\hspace{1cm}    If $\Psi \wedge \varphi$ is unsatisfiable, then $\Psi$
implies $\neg\varphi$:  \\
\hspace{1.5cm} remove $\varphi$ from all clauses in $\Phi$. \\
Loop until no literal has been removed\\
Return {\it satisfiable}.
\end{tabular}}
\end{center}
\vskip0.5\baselineskip

\noindent For testing whether a set of linear inequalities is satisfiable, we use a polynomial-time algorithm for linear program
feasibility, such a the ellipsoid method~\cite{Khachiyan};
it is well-known that 
this method can also be adapted to the situation
where some of the inequalities are strict, see e.g.~\cite{JonssonBaeckstroem}. 
Since the algorithm always
removes false literals, it is clear that if the algorithm returns \emph{unsatisfiable},
then $\Phi$ is indeed unsatisfiable. 
Suppose now that we are in the final step of the algorithm
and the procedure returns \emph{satisfiable}. 
Then for each negative literal $\varphi$ of $\Phi$ the set 
$\Psi \wedge \varphi$ has a solution $s_\varphi \colon V \to \mathbb Q$. 
Let $s$ be the mapping $s(x) \eqdef \max_{\varphi \text{ negative literal of } \Phi} s_\varphi(x)$. Clearly, $s$ satisfies $\Psi$
since the mappings $s_\varphi$ satisfy $\Psi$,
and since all literals in $\Psi$ are max-closed. 
We claim that $s$ satisfies all negative 
literals $\varphi = (a_1 x_1 + \cdots a_n x_n \succ c)$
in $\Phi$, too:
\begin{align*}
c \prec & \; a_1 s_\varphi(x_1) + \cdots + a_n s_\varphi(x_n) && \text{since $s_\varphi$ satisfies $\varphi$} \\
\leq & \; a_1 s(x_1) + \cdots + a_n s(x_n) && \text{since $a_1,\dots,a_n$ are positive.}
\end{align*}
Since every clause of~$\Phi$ either contains a negative literal or it is contained in~$\Psi$,
this shows that $s$ satisfies all constraints in $\Phi$. 

\section{Maximality of Max-closed Semilinear Relations}
\label{sect:maximal}
\begin{proposition}
Let $R \subseteq {\mathbb Q}^n$ an $n$-ary relation that is \emph{not} max-closed. Then
$\Csp(\Gamma_0,R)$ is $\NP$-hard. 
\end{proposition}
\begin{proof}
The proposition follows easily from~\cite[Theorems~6.5--6.6]{Ordered}, as
it is very short we offer here a direct argument.

Let $s,t \in R$ be such that $m=\max(s,t) \notin R$. Clearly the
finite relation $S\eqdef\{s,t,m\}$ is max-closed and semilinear, hence $R\cap S =
\{s,t\}$ is primitive positive definable in~$(\Q;\Gamma_0,R)$. Now, consider the
$3n$-ary relation
\[
T(x,y,z) \quad \Leftrightarrow \quad x,y,z\in R\cap S \; \wedge \; \max(x,y,z)=m
\]
which is also primitive positive definable in~$(\Q;\Gamma_0,R)$. Clearly
\[
T = (R\cap S)^3 \setminus \{(s,s,s), (t,t,t)\}
\]
and this gives a direct reduction of the $\NP$-complete problem of \emph{positive not-all-equal 3SAT}~\cite{GareyJohnson}
\[
\Csp\big(\{0,1\};\{0,1\}^3 \setminus \{(0,0,0),(1,1,1)\}\big)
\]
to~$\Csp(\Gamma_0,R)$.
\end{proof}


\end{document}